\documentclass[a4paper, 12pt]{article}

\usepackage[english]{babel}
\usepackage[utf8x]{inputenc}
\usepackage[T1]{fontenc}

\usepackage[top=1.3cm, bottom=2.0cm, outer=2.5cm, inner=2.5cm, heightrounded,
marginparwidth=1.5cm, marginparsep=0.6cm, margin=2cm]{geometry}

\usepackage{graphicx} 
\usepackage[colorlinks=False]{hyperref} 
\usepackage{xcolor}
\usepackage{mathtools}
\usepackage{amsmath}  
\usepackage{amsfonts} %
\usepackage{amssymb}  %
\usepackage{authblk}
\usepackage{cite}

\makeatletter
\newcommand\appendix@section[1]{%
  \refstepcounter{section}%
  \orig@section*{Appendix \@Alph\c@section: #1}%
  \addcontentsline{toc}{section}{Appendix \@Alph\c@section: #1}%
}
\let\orig@section\section
\g@addto@macro\appendix{\let\section\appendix@section}
\makeatother

\usepackage{caption}
\usepackage{float}

\def\nn{\nonumber \\ }

\newcommand{\pa}[1]{\left(#1 \right)}

\newcommand{\BR}[1]{\Biggl[#1 \Biggr]}

\newcommand{\ca}[1]{\mathcal{#1}}

\newcommand{\ar}[1]{\xrightarrow[#1]{}}

\newcommand{\fr}{\frac}

\def\la{{\lambda}}

 \def\d{{\delta}}
\def\sgn{{\text{sgn}}}

\def\vev#1{\langle\, #1 \, \rangle}
\def\ket#1{\mid \! #1\rangle}
\def\bra#1{\langle \, #1 \! \mid\! \ }
\def\braket#1#2{{\langle \, #1 \! \mid \! #2 \, \rangle}}
\def\Tr{{\rm Tr}}
\def\O{\mathcal{O}}
   
\def\be{\begin{equation}}
\def\ee{\end{equation}}

\title{Universality in asymptotic bounds and its saturation in $2$D CFT}
\author[1]{Diptarka Das\,}
\author[2]{Yuya Kusuki}
\author[3]{Sridip Pal}

\affil[1]{\small{Department of Physics, Indian Institute of Technology - Kanpur\\
Kanpur, UP - 208016, India.}}
\affil[2]{Center for Gravitational Physics,
Yukawa Institute for Theoretical Physics, Kyoto University, Kitashirakawa Oiwakecho, Sakyo-ku, Kyoto 606-8502, Japan.}
\affil[3]{School of Natural Sciences, Institute for Advanced Study,
Princeton, NJ - 08540, USA.}

 \date{}
  
\begin{document}
\begin{flushright}
YITP-20-142
\end{flushright}
\begingroup
\let\newpage\relax
\maketitle
\endgroup

\begin{abstract}
We study asymptotics of three point coefficients (light-light-heavy) and two point correlators in heavy states in unitary, compact $2$D CFTs. We prove an upper and lower bound on such quantities using numerically assisted Tauberian techniques. We obtain an optimal upper bound on the spectrum of operators appearing with fixed spin from the OPE of two identical scalars. While all the CFTs obey this bound, rational CFTs come close to saturating it. This mimics the scenario of bounds on asymptotic density of states and thereby pronounces an universal feature in asymptotics of 2D CFTs. Next, we clarify the role of smearing in interpreting the asymptotic results pertaining to considerations of eigenstate thermalization in 2D CFTs. In the context of light-light-heavy three point coefficients, we find that the order one number in the bound is sensitive to how close the light operators are from the $\frac{c}{32}$ threshold. In context of two point correlator in heavy state, we find the presence of an enigmatic regime which separates the $AdS_3$ thermal physics and the BTZ black hole physics. Furthermore, we present some new numerical results on the behaviour of spherical conformal block.
\end{abstract}

\clearpage

\section{Introduction \& Summary }

One of the fundamental pillars in physics is equivalence between microcanonical and canonical ensemble of systems with large degrees of freedom. In AdS-CFT, the incarnation of such equivalence is embodied in the thermal physics of black hole microstates, implying that a highly excited state in CFT should behave as if it is a thermal state. In particular, we expect there is a notion of ``chaotic'' CFT which can capture the ``chaotic'' and thermal features of black holes, for example, one should be able to see the ramp and plateau\cite{Cotler:2017jue} in the spectral form factor of such chaotic CFTs. While the lore in this context is that  a generic CFT with a twist gap (i.e. without any symmetry beyond Virasoro), and $c>1$ is ``chaotic'', there is no such known unitary modular invariant CFT partition function with a discrete spectrum. In spite of 
absence of an example, significant progress has been made in understanding the heavy excited states in  generic CFTs, with/without imposing the twist gap condition. A partial list of examples are Cardy formula \cite{cardy1986operator,HKS,baur, Ganguly:2019ksp,Pal:2019zzr, Mukhametzhanov:2020swe,Pal:2020wwd}, understanding physics related to ETH/black hole thermality \cite{KM,Lashkari:2017hwq,dattadaspal,Das:2017cnv,Cardy:2017qhl,Hikida:2018khg,Romero-Bermudez:2018dim,Brehm:2018ipf,Collier:2019weq,Fitzpatrick:2015zha,Fitzpatrick:2015foa,Fitzpatrick:2015dlt,Faulkner:2017hll,Anous:2019yku}, chaos at large central charge CFTs with sparse low lying spectrum \cite{Roberts:2014ifa}, identifying the irrational behavior in large $m$ minimal models \cite{Benjamin:2018kre}. 

A general theme of the above genre of work is to capture the high energy physics in a channel via vacuum block dominance in the dual channel. This is achieved by the virtue of crossing/modular symmetry and using some sort of gap condition (for example, in $2$D lightcone modular bootstrap \cite{Benjamin:2019stq,Kusuki:2018wpa}, one assumes twist gap, in higher D, twsit gap follows from unitarity and exploited in \cite{Komargodski:2012ek,Fitzpatrick:2012yx}). Although, the details might appear very different, in spirit, these methods mimic the derivation of Cardy formula. For example, in analyzing $2$ point correlator of light operators in heavy states in the large $c$ CFTs, one applies the monodromy method to obtain the vacuum block and then in a certain kinemetical configuration it is argued that the vaccum block dominates. Now to contrast this with the naive Cardy analysis, we note that the asymptotic density of states is similarly captured by the vaccum contribution in the dual channel. While in Cardy analysis, we obtain the vaccum contribution in the dual channel using modular invariance, in the study of correlators, this is achieved by modular covariance/crossing symmetry. Once we appreciate the conceptual parallelism between these two scenarios, we can use our recently understood refined knowledge about Cardy formula to understand the subtle features of general asymptotics in $2$D CFT. Recently, we have obtained a rigorous formulation of Cardy formula, where a smearing over an order one window is needed to make sense of the formula\cite{baur,Pal:2019zzr}. The basic thing we learnt was that the heavy states in the dual channel can actually have a cumulative efffect, thereby can control the averaging window \cite{baur}. This statement remains true even if there is a twist gap in the spectrum \cite{Pal:2019zzr}. Concretely, in context of asymptotic density of states in unitary modular invariant $2$D CFT, we have in $\Delta\to\infty$ limit \cite{baur, Ganguly:2019ksp, Mukhametzhanov:2020swe},
\begin{equation}
\begin{aligned}
(2\delta-1)\rho_0(\Delta) \leq \int d\Delta' \Theta_{(\Delta-\delta,\Delta+\delta)} \rho(\Delta') \leq \int \Theta_{[\Delta-\delta,\Delta+\delta]} \rho(\Delta') \leq (2\delta+1)\rho_0(\Delta) 
\end{aligned}
\end{equation}
where $\int d\Delta' \Theta_{(\Delta-\delta,\Delta+\delta)} \rho(\Delta')$ counts the states in the interval centered at $\Delta$ with width $2\delta$ excluding the end points and $\int d\Delta' \Theta_{[\Delta-\delta,\Delta+\delta]} \rho(\Delta')$ counts the states in the interval centered at $\Delta$ with width $2\delta$ including the end points. Here $\rho_0(\Delta)=(c/48\Delta^3)^{3/4}e^{2\pi\sqrt{\frac{c\Delta}{3}}}$ is the usual Cardy formula.  This inequality is saturated by CFTs with spectral gap of $1$. While the inequality is true for any unitary modular invariant CFTs, the rational CFTs almost saturate the inequality. Note the difference between the upper bound and the lower bound. Intuitively this comes about because a large $\rho_0(\Delta)$ number of states can accumulate at the end of the interval. This accounts for the $2\rho_0(\Delta)$ difference between the upper and the lower bound.  In a chaotic CFT, the difference between the upper and the lower bound is expected to be of the order of the spectral gap times $\rho_0(\Delta)$; thus, we expect (for a non-degenerate primary spectrum) the spectral gap is of the order of $\rho_0^{-1}$.

The purpose of the present work is to point out explicitly that the above features are universal, present in any asymptotic quantities in CFTs, not just restricted to the asymptotic density of states. While this is expected, there are technical hardles (which we will shed light on at the end of this section and throughout the paper) in proving so explicitly. In particular, we will be focussing on two point correlator in heavy states and light-light-heavy three point coefficients (often we will refine these asymptotics by restricting to primaries or primaries with fixed spin). The basic strategy is to derive an inequality of the following form
$$\widehat{\phi}_-(0) a_0(\Delta)\leq \int d\Delta' \Theta_{(\Delta-\delta,\Delta+\delta)} a(\Delta') \leq \int \Theta_{[\Delta-\delta,\Delta+\delta]} a(\Delta') \leq \widehat{\phi}_+(0)a_0(\Delta)\,,$$
where $a(\Delta)$ is the proxy for the quantity whose asymptotics we are trying to find and $a_0(\Delta)$ is a continuous approximation of the same. The functions $\widehat{\phi}_{\pm}$ has bounded support and their Fourier transform majorize and minorize  $\Theta_{[\Delta-\delta,\Delta+\delta]} $ and $\Theta_{(\Delta-\delta,\Delta+\delta)} $ respectively. The problem of finding the optimal value of $\widehat{\phi}_{\pm}(0)$ can be mapped to Beurling-Selberg problem and we can borrow results from \cite{Mukhametzhanov:2020swe}. To emphasize more on the universality of these qualitative features, let us point out one important feature in large central charge analysis in this genre. In context of Cardy formula at large central charge, \cite{HKS} pointed out the presence of an enigmatic regime i.e $\Delta\in(c/12,c/6)$. The $AdS_3$ thermal partition function is dominated by states below $c/12$ while the BTZ black hole is dominated by states above $c/6$. The states in the enigmatic regime never dominate the canonical ensemble. The similar feature is present in any large central charge asymptotics. For example, in calculation of two or $n$ point correlator in heavy state, the physics is captured by the vacuum block in dual channel \cite{Michel:2019vnk}. A natural way to calculate the vacuum block is monodromy method. The block being a continuous function of $\Delta$ can be calculated in one regime (say $\Delta<c/12)$ and can be analytically continued to the other regime $\Delta>c/12$. This has been repeatedly used in the literature \cite{Fitzpatrick:2015zha,Fitzpatrick:2015foa,Fitzpatrick:2015dlt,Faulkner:2017hll,Anous:2019yku}. The analogous statement in Cardy context is the fact the characters are smooth function of $h,\bar h$ and heavy density of states is captured by vacuum block. Thus we realize that the presence of the enigmatic regime is hidden in the assumption of dominance of vaccum block. Another unifying feature that we point out here is the commutativity of the limit $\Delta\to\infty$ (with $c$ fixed) followed by $c\to\infty $ limit with the $c\to\infty$ (with $\Delta/c$ fixed) followed by $\Delta/c\to\infty$ limit, present in the analysis of two point correlator, in the light-light-heavy three point coefficients and the asymptotic density of states\footnote{There can be subtlety when we refine these asymptotics only to heavy primaries.}.

One of the results resolves a puzzle mentioned in \cite{Faulkner:2017hll} .The puzzle is the fact how at large central charge the microcanonical answer for two point correlator in a heavy state ($\Delta/c$ fixed  and $c\to\infty$) of a CFT on a finite spatial length matches with the canonical answer for CFT on infinite line. Keeping in mind the analogy with Cardy formula just mentioned, a quick reader will observe a similar phenomenon happens at large central charge when one compares the microcanonical entropy of a CFT on a finite length with a canonical entropy of a CFT on infinite length. This puzzle gets resolved because the slogan ``microcanonical=canonical'' needs to be modified in this context as given in \eqref{holog} and not to be taken as granted. In particular as long the Euclidean time separation between two operators are less than the length of the spatial circle, microcanonical answer can not see the finite spatial length and we obtain canonical anaswer for CFT on infinite line. To be precise given a pure state with energy $\Delta$, the correlator can not see the finite length of spatial circle as long as as $t/L<1/2[3(\Delta/c-1/12)]^{-1/2}$, thus as we let $\Delta/c$ larger, the validity regime decreases, the less time survives the illusion of having a CFT on infinite length. If we could have let $\Delta/c\to 1/12$,the illusion would have remained forever. Nonetheless, we show that we need $\Delta/c>1/6$ for the result to be true, which means $t$ can not be bigger than $L$ i.e the illusion of being CFT on infinite length stays at most till $t=L$. 

Our results naturally fall into the scope of conformal bootstrap program \cite{Hellerman:2009bu,Collier:2016cls,Dyer:2017rul,Bae:2018qym, Alday:2019vdr,Lin:2019kpn,Cho:2017fzo,Bae:2017kcl,Afkhami-Jeddi:2019zci,Hartman:2019pcd,Benjamin:2019stq,Benjamin:2020swg,Benjamin:2020zbs,Afkhami-Jeddi:2020hde,Afkhami-Jeddi:2020ezh,Kaidi:2020ecu}. We hope the techniques/functions used here would be useful in the broader context of this program, especially for the studies related to extremal functionals and dispersive sum rules \cite{Paulos:2019gtx,Mazac:2018mdx,Mazac:2018ycv,Mazac:2019shk,Caron-Huot:2020adz} and its connection to analyticity in replica correlator \cite{Shyani:2020oja}.  \\

Concretely, our findings are summarized below (we have not assumed presence of twist gap in CFT spectra anywhere in the paper and the asymptotic analysis specific to primaries are only valid for $c>1$ and the CFT is assumed to be unitary and compact)
\begin{enumerate}
\item We analyze the fixed spin weighted asymptotics of light light heavy three point coefficients (refined to heavy primaries, with finite $c>1$). We define (where the $\pm$ signifies whether the end points of the interval are included or not)
$$\mathcal{A}_{j\pm}\equiv \int_{\Delta-\delta}^{\Delta+\delta} d\Delta^\prime\ |f_{\mathcal{O}\mathcal{O}\Delta'(j)}|^2$$ and we find that in $\Delta\to\infty$ limit
\begin{equation}
\begin{aligned}\label{mainineq}
\frac{2(\delta-2)}{1+\delta_{j,0}}\frac{1}{H^2}\mathcal{A}_{0}(\Delta) \leq \mathcal{A}_{j-} \leq\mathcal{A}_{j+}\leq \frac{2(\delta+2)}{1+\delta_{j,0}}\frac{1}{H^2} \mathcal{A}_{0}(\Delta) 
\end{aligned}
\end{equation}
where we have
$$\mathcal{A}_0(\Delta)\equiv\left[\frac{1}{4}  \left(\frac{c-1}{24}\right)^{1/4-\nu/2} \left(\frac{\Delta}{2}\right)^{\nu/2-5/4}e^{\pi\sqrt{\frac{(c-1)\Delta}{3}}}\right]$$
is the naive formula obtained by assuming dominance of vacuum block. The quantity $\nu=8\Delta_{\mathcal{O}}-c/2$ is determined by the dimension of external operators ($\Delta_{\mathcal{O}}$) and central charge $c$. The quantitiy, $H$ is an order one number that depends on $c$ and $\Delta_{\mathcal{O}}$. Here $|f_{\mathcal{O}\mathcal{O}\Delta'}|^2$ includes the factor of $16^{\Delta}$ in its definition, coming from the normalization condition and operator insertion point.

$\bullet$ Two features are worth to be pointed out. The number $H$ depends on how close the external operators $\mathcal{O}$ is to $\frac{c-1}{32}$ or $\tfrac{c+5}{32}$. In particular, if the difference is an order one number, then in the $c\to \infty$ limit (note that this limit is different from the limit where $\Delta/c$ is finite and $c\to\infty$, here we first take $\Delta\to\infty$, keeping $c$ fixed and then take $c\to\infty$), $H$ becomes $1$ and we recover the results obtained earlier in the literature . For a finite but large $c$, the value of $H$ stays close to $1$, but not quite $1$. A good approximation can be obtained from the formula \cite{Cardona:2020cfy}(further studied by \cite{Das:2020fhs}) 
\begin{equation}
H(h,q) = 1 - \fr{\pa{c+1-32h_{\ca{O}}}\pa{c+5-32h_{\ca{O}}}}{4(c-1)}+O(1/c^2)\,,\ \text{where}\ h_{\ca{O}}-c/32\simeq O(1)
\end{equation}

 When the difference $h_{\ca{O}}-c/32$ is large (or in fact proportional to $c$), numerically we have observed $H$ saturates to a value farther away from $1$. Unfortunately, we don't have any analytical control in this regime. The most conservative statement that we can make is the leading exponential in $\mathcal{A}_0(\Delta)$ stays same. A braver statement (once again, on numerical back up) would be that it would saturate to a finite value (may be far from $1$), and such a saturation indicates that we can trust the polynomial suppression in the above expression for $\mathcal{A}_0(\Delta)$ as well. The analytical results of this sections are made based on some assumptions which we back up by numerical study in \S~\ref{Sec:vir}. 

$\bullet$ The Bound on the spectral gap for fixed spin primary operators appearing in the OPE of two identical scalar operators is $4$. To prove the optimality, consider 3 copies of Ising model, organize the operators with respect to the diagonal Virasoro, since $c>1$, we have infinite number of primaries. Now consider the operator $\mathcal{O}=\epsilon\otimes\epsilon\otimes\epsilon$ and 4 point correlator of them. The primary operators with fixed spin (primary under the diagonal Virasoro) appearing in the OPE has a spectral gap of $4$ (here the gap in $h$ is $2$, the gap in $\bar h$ is $2$, thus in the fixed spin the gap in $\Delta$ is $4$). Another possible optimal construction is to consider tensoring chiral Monster with its antichiral avatar and consider $4$ point function of a scalar operator. Also, given a fixed spin spectrum of gap $4$, the inequalities \eqref{mainineq} are saturated asymptotically in exactly same manner as in \cite{Mukhametzhanov:2020swe}. And all the rational CFTs come close to saturation because they have a regularly spaced spectra. Further statements abour spectral gap in OPE channel can be found in \cite{cardy1986operator,HKS,baur, Ganguly:2019ksp, Mukhametzhanov:2020swe}. 

The above set of analysis involving $4$ point correlator is explored in \S~\ref{Sec:llh}. The analysis for all the operators (not specific to three point coefficient for primary operators) is worked out in \eqref{allops}. The analysis for all the operators are done with respect to a single parameter $\Delta\to \infty$. One can generalize this to an refined counting with respect to $h,\bar h\to\infty$. While doing this, it is more convenient to distill the counting further and restrict to primaries (for $c>1$ CFTs, we have inifnite number of primaries). This is done in \eqref{primary}. Finally, we have the most refined count i.e.\!~primaries at fixed spin.

\item The $n$ point correlator in heavy state in large c CFT behaves like a thermal state \cite{Faulkner:2017hll,Anous:2019yku}. In large central charge the distinction between heavy primary and heavy descendants are not important, so we focus on analysis for all the operators. We point out and stress the presence of an enigmatic regime $\Delta\in (c/12,c/6)$. In particular, we show that these genre of statements are true for $\Delta>c/6$ only. We analyze the $2$ point correlator in some detail. Similar ideas can be extended to higher point correlators. This is explored in \S~\ref{Sec:npoint}.

If we parametrize $\Delta=c\left(\frac{1}{12}+\epsilon\right)$ , then in $c\to\infty$ limit, for $t\leq\beta=\frac{\pi}{\sqrt{3\epsilon}}$ (where $t$ is the Euclidean time) and $\Delta\geq c/6$ we have 
\begin{equation}
\begin{aligned}\label{holog}
&\ \left(2\delta-\left(1-\frac{\beta^2}{4\pi^2}\right)^{-1/2}+\ell\right) e^{-\beta\delta}\rho_0(\Delta)\left(\frac{2\pi}{\beta}\right)^{2\Delta_{\mathcal{O}}}\left[\sin\left(\frac{\pi t}{\beta}\right)\right]^{-2\Delta_{\mathcal{O}}}\\
&\leq \mathcal{B}_{-}\leq\mathcal{B}_+ \\
&\leq \left(2\delta+\left(1-\frac{\beta^2}{4\pi^2}\right)^{-1/2}+\ell\right)e^{\beta\delta}\rho_0(\Delta)\left(\frac{2\pi}{\beta}\right)^{2\Delta_{\mathcal{O}}}\left[\sin\left(\frac{\pi t}{\beta}\right)\right]^{-2\Delta_{\mathcal{O}}}
\end{aligned}
\end{equation}
where $$\rho_0(\Delta)=\left(\frac{c}{48(\Delta-c/12)^3}\right)^{3/4}\exp\left[2\pi\sqrt{\frac{c(\Delta-c/12)}{3}}\right]$$ and
$$\mathcal{B}_{\pm}\equiv \int_{\Delta-\delta}^{\Delta+\delta} d\Delta'\langle \Delta'| \mathcal{O}(t,0)\mathcal{O}(0,0) |\Delta' \rangle\,,$$
here $\pm$ signifies whether we are including or excluding the end points of the interval, centered at $\Delta$ and $\ell$ is an order one positive number coming from the contribution of sparse low lying states. If one assumes that the identity is the only state below the $\frac{c-1}{12}$ threshold, then $\ell=0$. Loosely speaking, we prove that the average behaviour of two point correlator in heavy state in large central charge CFT on a finite spatial length ($2\pi$) mimics that of a thermal CFT on infinite line. Note the result is valid as long as $t\leq \beta$ and the maximum allowed value of $\beta$ is $2\pi$ since $\epsilon>1/12$. Restoring the length of the spatial circle $L$, we find the result is valid as long as $t<L$ i.e as long as the Euclidean time seperation is less than spatial length, microcanonical answer sees the canonical answer for CFT on infinite line.
We note the presenece of enigmatic regime $\Delta\in (c/12,c/6)$, this never dominates the canonical ensemble. This is reminiscent of Cardy formula for large central charge as done in HKS. The presence of enigmatic regime is in fact the reason for having $t<L$ bound. 

$\bullet$ For finite $c$, $\Delta\to\infty$ the above inequality becomes
\begin{equation}
\begin{aligned}\label{finitec}
&\ \left(2\delta-1\right)\rho_0(\Delta)\left(\frac{2\pi}{\beta}\right)^{2\Delta_{\mathcal{O}}}\left[\sin\left(\frac{\pi t}{\beta}\right)\right]^{-2\Delta_{\mathcal{O}}}\\
&\leq \mathcal{B}_{-}\leq\mathcal{B}_+ \\
&\leq \left(2\delta+1\right)\rho_0(\Delta)\left(\frac{2\pi}{\beta}\right)^{2\Delta_{\mathcal{O}}}\left[\sin\left(\frac{\pi t}{\beta}\right)\right]^{-2\Delta_{\mathcal{O}}}
\end{aligned}
\end{equation}
It is to be understood as $\Delta\to\infty$ limit result such that $\beta=\pi\sqrt{\frac{c}{3\Delta}}$ goes to $0$ and $t\to 0$ with $t/\beta$ fixed at some finite number. 

$\bullet$ \textbf{Commutativity of the limit:} We remark that starting from \eqref{holog}, if one takes $\epsilon\to\infty$ limit such that $\beta\to 0$ and $t/\beta$ is kept fixed (i.e $t\to 0$ as well), one arrives at \eqref{finitec}. In $\beta\to 0$ limit, the contribution from low lying states other than the vacuum gets suppressed, hence $\ell\to 0$. This indicates the commutativity of the following two limits 
\begin{equation}
\lim_{\kappa\to\infty}\lim_{\underset{\frac{\Delta}{c}=\kappa\ \text{fixed}}{c\to\infty}} \left[\cdots\right]= \lim_{c\to\infty}\lim_{\underset{c=\text{fixed}}{\Delta\to\infty}} \left[\cdots\right]
\end{equation}

This feature is present in analysis of asymptotic density of states in $2$D CFT.
\end{enumerate}

As a prerequisite for the analysis of light light heavy three point coefficients, we studied asymptotics of  Virasoro block and provide numerical results. These are described in \S~\ref{Sec:vir} and can be read independently of the rest of the sections.\\ 

At this point, an inquisitve reader might wonder about the absence of large $c$ analysis in the light-light-heavy case and absence of primary specific analysis for the $2$ point correlator in finite $c$. The light-light-heavy case can easily be done for large $c$, we write down here the analogue of \eqref{allops} for future reference 
\begin{equation}
\begin{aligned}
&\left(\delta-\left(1-\frac{\beta^2}{4\pi^2}\right)^{-1/2}\right)\left[16^{-c/12}\right]e^{-\beta\delta/2}\left(\frac{c}{12}\right)^{\frac{1}{4}-\frac{\nu }{2}} \left(\Delta/c-1/12\right) ^{\frac{\nu }{2}-\frac{3}{4}}e^{\pi \sqrt{\frac{c(\Delta-c/12)}{3}}} \\
&\leq \int_{0}^{\infty} d\Delta^\prime\  \Theta\left((\Delta-\delta,\Delta+\delta)\right)a(\Delta^\prime)\sigma(\Delta^\prime) \\
&\leq \int_{0}^{\infty} d\Delta^\prime\  \Theta\left([\Delta-\delta,\Delta+\delta]\right)a(\Delta^\prime)\sigma(\Delta^\prime) \\
&\leq\left(\delta+\left(1-\frac{\beta^2}{4\pi^2}\right)^{-1/2}\right)\left[16^{-c/12}\right]e^{\beta\delta/2}\left(\frac{c}{12}\right)^{\frac{1}{4}-\frac{\nu }{2}}\left(\Delta/c-1/12\right) ^{\frac{\nu }{2}-\frac{3}{4}}e^{\pi \sqrt{\frac{c(\Delta-c/12)}{3}}}
\end{aligned}
\end{equation}
where $\Delta=c\left(1/12+\epsilon\right)$ and we let $c\to\infty$, keeping $\beta=\frac{\pi}{\sqrt{3\epsilon}}$ fixed. The qualitative features including the presence of enigmatic regime are universal as advertised. The primary specific analysis for the $2$ point correlator in finite $c$ is tricky because what we need is an analog of half of a torus 2 point block, whose analytic form is not known. Also it is unclear whether universal results can be expected beyond the $t\to 0$ limit. The intuition partly comes from the result that primary states are atypical in a thermal ensemble for finite $c$ \cite{Dymarsky:2018lhf,Datta:2019jeo}. We leave it for future exploration. \\

$\bullet$ \textbf{Technical remarks:}
The pivotal technical tool that we use is Tauberian tehniques coupled with some numerical studies of Virasoro block. In context of CFT, Tauberian formalism was first introduced in \cite{Pappadopulo:2012jk,Qiao:2017xif}, generalized in \cite{Mukhametzhanov:2018zja}, mentioned in appendix C of \cite{dattadaspal}, culminating to refined understanding of Cardy formula in \cite{baur, Ganguly:2019ksp, Pal:2019zzr, Pal:2019yhz, Mukhametzhanov:2020swe}. The main technical obstacle behind generalization to OPE coefficients at finite central charge is that the conformal blocks are not known in a closed form unlike the Virasoro characters in context of Cardy formula. While in the analysis of Cardy formula, the presence of closed form of characters and nice modular property of Dedekind eta helped us to write the asymptotic density of primaries as an inverse Laplace of reduced partition function in $\beta\to 0$ limit, making the Tauberian analysis possible\cite{Mukhametzhanov:2020swe}. Thus while dealing with correlators, one needs further assumptions to proceed. This is the route we take by backing up the assumptions with numerical estimations. Similarly for the analysis of two point function, we focus on large central charge analysis, to get control over the correlator. At large central charge the distinction between primaries and all the operators are not that important. Thus  we also restrict our analysis to all operators. Nonetheless, a further complication arises due to the fact that the under modular transformation, operator insertion points transform, so one needs to take that into account as well. For finite $c$, it is much more challenging to restrict the analysis to primaries and we have not attempted to do so in this paper. Nonetheless we believe that qualitative features/moral lessons stay same. We further remark that unlike $HLH$ case, here we do not have the issue of negative terms in $q$ expansion of relevant quantities. So the results proven here do not require stringent conditions appearing in \cite{Pal:2019yhz}.

Finally, it is tempting to suspect that some of the analysis especially the light-light-heavy OPE asymptotics might be obtained in higher dimensional CFT using crossing symmetry and similar Tauberian analysis. Again the main technical obstacle is the presence of conformal blocks. Even if we know the blocks explicitly, there is no clean way of writing a reduced partition function as a Laplace transformation of weighted OPE coefficients. Nonetheless, it is tempting to conjecture that the such bounds can be obtained and ope of generalized free fields would saturate it \cite{sb} including the bound on the spectral gap.

\section{Tauberian theorem(s) for Asymptotics of Light-Light-Heavy }\label{Sec:llh}

\def\vev#1{\langle\, #1 \, \rangle}
\def\ket#1{\mid \! #1\rangle}
\def\bra#1{\langle \, #1 \! \mid\! \ }
\def\braket#1#2{{\langle \, #1 \! \mid \! #2 \, \rangle}}
\def\Tr{{\rm Tr}}
\def\O{\mathcal{O}}

We start with the four point function of identical scalar primaries on the sphere, 
${\cal F}(z,\bar{z} ) = \vev{  \O(0)\O(z,\bar{z} )\O(1)\O(\infty) }$, which has a standard decomposition into Virasoro conformal blocks on the sphere and satisfies the crossing equation non-trivially (the individual blocks are not crossing symmetric):
\begin{equation} 
{\cal F}( z , \bar{z} ) = {\cal F}( 1- z,1-\bar{z} ). \label{crossing1} 
\end{equation}
There is no closed form known for the Virasoro blocks, however each term in the cross-ratio expansion can be determined using conformal Ward identities. A better expansion than the expansion in $z$ turns out to be an expansion in $q = e^{\pi i \tau}$, where $\tau = i K(1-z)/K(z)$, and $K(z)$ being the elliptic integral of the first kind. Such an expansion originates from the elliptic representation of the conformal blocks, in the pillow geometry \cite{msz} which is $\mathbb{P}^1 \equiv \mathbb{T}^2/ {\mathbb{Z}}_2$  which is conformally equivalent to the sphere. Crucially, on the pillow the symmetry of the correlator under $z \rightarrow 1 -z$, \eqref{crossing1} becomes, $\tau \rightarrow -1/\tau$.

\subsection{Analysis for all the operators} 

For pillow, we will be working with following function $g$
\begin{align}\label{relate}
g(q , \bar q)= {\cal F}(z, \bar z) \Lambda(z)^{-1}\Lambda(\bar z)^{-1} 
\end{align}
Here $\Lambda(z) \equiv \vartheta_3(q)^{{c\over 2}-8\Delta_\O}(z(1-z))^{{c\over 24}-\Delta_\O}$, which contains contributions from the conformal factors at the operator insertions and the Weyl anomaly factor arising due to change of conformal frame. $g(q,\bar q)$ is the regularized correlator on the pillow which is defined as
\begin{align}
g(q,\bar q) &\equiv \vev{  \O(0)\O(\pi)\O(\pi(\tau+1))\O(\pi\tau) } _{\mathbb{P}^1}\nn \\
&= \vev{ \psi | q^{L_0-c/24} \bar{q}^{\bar{L}_0-c/24}| \psi  }, \label{g-def}
\end{align}
with $\ket{\psi}= \ket{\O(\pi)\O(0)}_{\mathbb{P}^1}$.
The decomposition of the pillow correlator takes the following form (with $\tau = \tfrac{i \beta}{2\pi }$ ) in terms of the elliptic conformal blocks: 
\begin{equation}\label{decom}
g(\beta) = \sum_{h,\bar{h}} f^2_{\O\O\O_{h,\bar{h}}} {\cal V}_h(q) \tilde{\cal V}_{\bar{h} }(\bar{q} ).
\end{equation}
where, 
\begin{equation}\label{eq:Vh}
\begin{aligned}
{\cal V}_h(q) &=  G(h,q) H(h,q), \\
G(h,q)&\equiv (16 q )^{h  - \frac{c}{24} } \prod_{k=1}^\infty ( 1- q^{2k } )^{-\frac{1}{2} } \pa{= (16 q )^{h  - \frac{c-1}{24} }  \theta_3 (q)^{-\frac{1}{2}} (z(1-z))^{-\frac{1}{24}}  }.
\end{aligned}
\end{equation}
Note that the relation to the usual convention for the conformal block $\ca{F}(h|z)$ with its internal dimension $h$ is given by
\begin{equation}\label{eq:usualF}
\ca{F}(h|z) = \Lambda(z) {\cal V}_h(q).
\end{equation}
In the above expression $H(h,q) = \sum_{n} c_n q^n$ is the Zamolodchikov block, for which a recursion formula exists \cite{zamu}. 
For a unitary theory it is guaranteed that the above $q$ expansion is a one with positive coefficients, hence we can expand:
\begin{align} \label{decomp}
g(\beta) &=  \sum_{\Delta} a(\Delta) e^{-\tfrac{\beta}{2} (\Delta - \tfrac{c}{12}) }
\end{align}
In the above we have the sum over all operators (including the descendents exchanged in the correlator channel). The coefficients, $a(\Delta) \geq 0$ due to unitarity \cite{msz}.  
The exact consequence of \eqref{crossing1} in the $\beta$ variable for the pillow correlator, is given by,
\begin{align}\label{modpill}
g(\beta)&=\left(\frac{2\pi}{\beta}\right)^{-\frac{c}{2}+8\Delta_\O}g\left(\frac{4\pi^2}{\beta}\right).
\end{align}
This modular property has already been utilized in \cite{Das:2017cnv, kusuki1} to bootstrap $f^2_{\O\O\O_{h,\bar{h}}}$ in the asymptotic heavy channel limit, which has also been reproduced through the Virasoro fusion kernel as carried out in \cite{Collier:2018exn}. 

Next we define a discrete distribution of operators, 
\begin{align}
\sigma(\Delta^\prime)= \sum_{\Delta}\delta\left(\Delta-\Delta^\prime\right),
\end{align}
and we will be estimating the following weighted spectral density : 
\begin{align}
\mathcal{A}\equiv \int_{\Delta-\delta}^{\Delta+\delta} d\Delta^\prime a(\Delta^\prime)\sigma(\Delta^\prime).
\end{align}
This can be interpreted to be proportional to the number of states that is being exchanged in the $q$-channel of the pillow correlator, with $L_0 + \bar{L}_0$ eigenvalue within the window specified by, $\Delta - \delta$ to $\Delta + \delta$. Depending on whether we include the end points while counting the number of operators, we should define $\mathcal{A}_+$ and $\mathcal{A}_{-}$, where $\pm$ means inclusion and exclusion respetively. Thus the lower bound that follows is actually a lower bound on $\mathcal{A}_-$ and the upper bound is an upper bound on $\mathcal{A}_+$. Futhermore, by definition, we have $\mathcal{A}_+\geq\mathcal{A}_-$. We will mostly suppress the $\pm$ index in $\mathcal{A}_{\pm}$ for brevity in what follows.

We start with band-limited functions $\Phi_{\pm}$ such that it bounds the indicator function: 
\begin{align}
\Phi_-(\Delta^\prime)\leq \Theta ( \Delta' \in [ \Delta- \delta , \Delta + \delta]) \leq \Phi_+(\Delta^\prime).
\end{align}
From the above it naturally follows (after multiplying throughout by $a(\Delta')$ and integrating over the all $\Delta'$) that,
\begin{equation}
\begin{aligned}
&e^{\tfrac{\beta}{2}(\Delta-\delta)}\int d\Delta^\prime a(\Delta^\prime)\sigma(\Delta^\prime)\Phi_-(\Delta^\prime)e^{-\tfrac{\beta}{2}\Delta^\prime} \leq {\cal A} \leq
e^{\tfrac{\beta}{2}(\Delta+\delta)}\int d\Delta^\prime a(\Delta^\prime)\sigma(\Delta^\prime)\Phi_+(\Delta^\prime)e^{-\tfrac{\beta}{2}\Delta^\prime} .
\end{aligned}
\end{equation}
Next we use the Fourier transformation of the band-limited functions,
$
\Phi_{\pm}(\Delta)= \int_{-\infty}^{\infty}dt e^{-i \Delta t}\widehat{\phi}_{\pm}(t)
$
to write the above inequality as,
\begin{align}
e^{\tfrac{\beta}{2} ( \Delta - \delta ) } \int dt\ \widehat{\phi}_-(t) e^{- \tfrac{(\beta + 2i t)c }{24}  }  g(\beta + 2 i t ) \leq {\cal A} \leq 
e^{\tfrac{ \beta}{2} ( \Delta + \delta ) } \int dt\ \widehat{\phi}_+(t) e^{- \tfrac{(\beta + 2i t)c}{24}  } g(\beta + 2i t ). 
\end{align}

Thus we need to evaluate the integrals of the form
$$I_{\pm}\equiv \int dt\ \widehat{\phi}_{\pm}(t) e^{- \tfrac{(\beta + 2i t) c}{24} } g(\beta + 2i t )$$
Using the property \eqref{modpill} we arrive at,
\begin{equation}
I_{\pm}= \int dt\ \widehat{\phi}_{\pm}(t) \left( \frac{ 2\pi}{ \beta + 2 i t } \right)^{8 \Delta_\O - \tfrac{c}{2} } e^{- \tfrac{(\beta + 2i t)c }{24} }  g\left(\tfrac{4\pi^2}{\beta + 2 i t} \right)
\end{equation}
 
 We separate the $g$ into light and heavy contributions as follows:
 \begin{equation}
 g_L(\beta)=\sum_{\Delta<c/12}a(\Delta)e^{-\beta(\Delta-c/12)}\,,\  g_H(\beta)=\sum_{\Delta>c/12}a(\Delta)e^{-\beta(\Delta-c/12)}\,,
 \end{equation}
 leading to 
 \begin{equation}
 \begin{aligned}
&e^{\tfrac{\beta}{2} ( \Delta - \delta ) }\left( \int dt\  \widehat{\phi}_-(t)   e^{- \tfrac{\beta(t) c }{24} } \left( \frac{ 2\pi}{ \beta(t)  } \right)^{\nu} g_L\left(\tfrac{4\pi^2}{\beta(t) } \right)
-\int dt\ \bigg| \widehat{\phi}_-(t) e^{- \tfrac{\beta(t) c }{24} } \left( \frac{ 2\pi}{ \beta(t)  } \right)^{\nu}  g'_H\left(\tfrac{4\pi^2}{\beta(t) } \right)\bigg|\right)\\
& \leq {\cal A} \leq \\
&e^{\tfrac{\beta}{2} ( \Delta + \delta ) } \left(\int dt\ \widehat{\phi}_+(t)   e^{- \tfrac{\beta(t) c }{24} } \left( \frac{ 2\pi}{ \beta(t)  } \right)^{\nu} g_L\left(\tfrac{4\pi^2}{\beta(t) } \right)+\int dt\ \bigg| \widehat{\phi}_+(t)  e^{- \tfrac{\beta(t) c }{24} }  \left( \frac{ 2\pi}{ \beta(t)  } \right)^{\nu}g_H\left(\tfrac{4\pi^2}{\beta(t) } \right)\bigg|\right)
\end{aligned}
 \end{equation}
 where $\beta(t)\equiv \beta+2\imath t$ and $\nu=8\Delta_{\mathcal{O}}-c/2$.\\
 
 Now for the light part, in $\beta\to 0$ limit, we have
\begin{equation}\label{lightpart}
g_L\left(\tfrac{4\pi^2}{\beta(t) } \right)\underset{\beta\to0}{=}\left[16^{-c/12}\right] e^{\frac{\pi^2c}{6\beta(t)}}\,.$$  $t=0$ is a saddle and the integral evaluates to $$\int dt\ \phi_{\pm}(t)   e^{- \tfrac{\beta(t) c }{24} } \left( \frac{ 2\pi}{ \beta(t)  } \right)^{\nu} g_L\left(\tfrac{4\pi^2}{\beta(t) } \right)\underset{\beta\to0}{=} \left[16^{-c/12}\right] \widehat{\phi}_{\pm}(0)\sqrt{\frac{3}{2\pi c}}\ \beta^{3/2}\left(\frac{ 2\pi}{ \beta} \right)^{\nu} e^{\frac{\pi^2c}{6\beta}}
\end{equation}
 
For the heavy part, we use the bandlimited nature of $\widehat{\phi}_{\pm}$ to write  
$$\int_{-\Lambda}^{\Lambda} dt\ \bigg| \widehat{\phi}_+(t)  e^{- \tfrac{\beta(t) c }{24} }  \left( \frac{ 2\pi}{ \beta(t)  } \right)^{\nu}g_H\left(\tfrac{4\pi^2}{\beta(t) } \right)\bigg|\leq \int_{-\Lambda}^{\Lambda} dt\ \bigg| \widehat{\phi}_+(t) \bigg| \left( \frac{ 2\pi}{ \sqrt{\beta^2+4t^2} } \right)^{\nu}g_H\left(\tfrac{4\pi^2\beta}{\beta^2+4t^2} \right)\,,$$
and the integral on the right hand side is dominated by $t=\Lambda$ (since we have $g_{H}\ni e^{-\tfrac{2\pi^2\beta}{\beta^2+4t^2} \left(\Delta-c/12\right)}$ ) and we have  
\begin{equation}\label{heavypart}
\int_{-\Lambda}^{\Lambda} dt\ \bigg| \widehat{\phi}_{\pm}(t) \bigg| \left( \frac{ 2\pi}{ \sqrt{\beta^2+4t^2} } \right)^{\nu}g_H\left(\tfrac{4\pi^2\beta}{\beta^2+4t^2} \right) \simeq  O\left(e^{\frac{\Lambda^2 c}{6\beta}} \beta^{-\nu+2} \right)
\end{equation}
where we have used the $ \widehat{\phi}_{\pm}(t) \sim O(\Lambda-t)$. Combining the light part and the heavy part, we obtain
\begin{equation}
\begin{aligned}
&2\pi\widehat{\phi}_{-}(0)\left[16^{-c/12}\right]\sqrt{\frac{3}{8\pi^3 c}}\ \beta^{3/2}\left(\frac{ 2\pi}{ \beta} \right)^{\nu} e^{\frac{\pi^2c}{6\beta}+\frac{\beta}{2}\Delta}+O\left(e^{\frac{\Lambda^2c}{6\beta}+\frac{\beta}{2}\Delta} \beta^{-\nu+2} \right)\\
&\leq \int_{\Delta-\delta}^{\Delta+\delta} d\Delta^\prime a(\Delta^\prime)\sigma(\Delta^\prime)\leq\\
&2\pi\widehat{\phi}_{+}(0)\left[16^{-c/12}\right]\sqrt{\frac{3}{8\pi^3 c}}\ \beta^{3/2}\left(\frac{ 2\pi}{ \beta} \right)^{\nu} e^{\frac{\pi^2c}{6\beta}+\frac{\beta}{2}\Delta}+O\left(e^{\frac{\Lambda^2c}{6\beta}+\frac{\beta}{2}\Delta} \beta^{-\nu+2} \right)
\end{aligned}
\end{equation}

Now to suppress the heavy part, the maximum value of $\Lambda$ that can be chosen is $\Lambda=\pi$. The Beurling-Selberg function corresponding to $\Lambda=\pi$ yields

$$2\pi\widehat{\phi}_{\pm}(0)=2(\delta\pm1)\,,$$
and minimizing the first term over $\beta$ gives the thermodynamic relation between $\beta$ and $\Delta$ as $\beta=\pi\sqrt{\frac{c}{3\Delta}}$. Thus we have in $\Delta\to\infty $ limit
\begin{equation}\label{allops}
\begin{aligned}
&(\delta-1)\left[16^{-c/12}\right]\left(\frac{c}{12}\right)^{\frac{1}{4}-\frac{\nu }{2}} \Delta ^{\frac{\nu }{2}-\frac{3}{4}}e^{\pi \sqrt{\frac{c\Delta}{3}}} \\
&\leq \int_{0}^{\infty} d\Delta^\prime\  \Theta\left((\Delta-\delta,\Delta+\delta)\right)a(\Delta^\prime)\sigma(\Delta^\prime) \\
&\leq \int_{0}^{\infty} d\Delta^\prime\  \Theta\left([\Delta-\delta,\Delta+\delta]\right)a(\Delta^\prime)\sigma(\Delta^\prime) \\
&\leq(\delta+1)\left[16^{-c/12}\right]\left(\frac{c}{12}\right)^{\frac{1}{4}-\frac{\nu }{2}} \Delta ^{\frac{\nu }{2}-\frac{3}{4}}e^{\pi \sqrt{\frac{c\Delta}{3}}}
\end{aligned}
\end{equation}
where $\nu=8\Delta_{\mathcal{O}}-c/2$ and we have explicitly denoted two cases: inclusion and exclusion of end points while counting the contribution.

The lower bound implies that asymptotically we have a maximal gap of $2$ in spectrum of operators appearing in the OPE of Identical scalars in $q$-channel. This is in fact the optimal bound as can be seen from 4 point correlator of $\epsilon$ operator ($h=\bar h=1/2$) in $2$-D Ising CFT ($c=1/2$):
$$\mathcal{F}_{\epsilon\epsilon\epsilon\epsilon}(z,\bar{z})=\frac{\left(1-z+z^2\right)\left(1-\bar z+\bar z^2\right)}{z\bar{z}(1-z)(1-\bar z)}=\mathcal{F}_{\epsilon\epsilon\epsilon\epsilon}(1-z,1-\bar{z})\,,$$
where $z=z(\tau),\bar{z}=\bar{z}(\tau)$ should be understood as a function of $\tau$, then one transforms it to the pillow correlator $g_{\epsilon\epsilon\epsilon\epsilon}(\beta)$ including appropriate Weyl anomaly factor following \eqref{relate}.

\subsection{Analysis for primaries on $(h',\bar h')$ plane} 
In this section, we make the analysis from earlier section sensitive to primaries. We also introduce separate left moving and right moving temperatures $\beta_L$ and $\bar \beta_R$ with an aim to make the analysis sensitive to $h$ and $\bar h$. On the plane we can reach asymptotic region in various ways as explained in \cite{Pal:2019zzr}. Here we will focus on reaching asymptotics along the line $h=\bar h+J$ where $J$ is an order one number. This specific choice will simplify our life.\\
%

We introduce the following quantity
\begin{align}
p(\beta_L,\beta_R)= \sqrt{\eta \left(\beta_L\right)\eta\left(\beta_R\right)} g(\beta_L,\beta_R) 
\end{align}
which has following expansion:
\begin{equation}
\begin{aligned}\label{pexp}
p(\beta_L,\beta_R)&=\sum_{h', \bar{h}'}f^{2}_{\mathcal{O}\mathcal{O}O'}e^{-\frac{\beta_L}{2}\left(h'-\frac{C}{24}\right)-\frac{\beta_R}{2}\left(\bar h'-\frac{C}{24}\right)}H(h',q_L)H(\bar{h}',q_R)\\
\end{aligned}
\end{equation}
where $C=c-1$ and $f^{2}_{\mathcal{O}\mathcal{O}O'}$ includes a factor of $16^{\Delta-c/12}$.
We define an auxiliary function $p_{\text{aux}}(\beta,\omega)$
\begin{equation}
p_{\text{aux}}(\beta_L,\beta_R)= \sum_{h', \bar{h}'}f^{2}_{\mathcal{O}\mathcal{O}O'}e^{-\frac{\beta_L}{2}\left(h'-\frac{C}{24}\right)-\frac{\beta_R}{2}\left(\bar h'-\frac{C}{24}\right)}\,,
\end{equation}
whose role will be important very shortly.

Using modular transformation, we deduce that
\begin{equation}\label{modtrafo}
p(\beta_L,\beta_R)\underset{\beta_{L,R}\to0}{\simeq}\left(\frac{4\pi^2}{\beta_L\beta_R}\right)^{\nu/2+1/4}\exp\left[\frac{\pi^2(c-1)}{12\beta_L}+\frac{\pi^2(c-1)}{12\beta_R}\right]\,.
\end{equation}

Our objective is to estimate the following quantity
\begin{equation}
\begin{aligned}
&\mathcal{A}\equiv \int_{h-\delta}^{h+\delta}\int_{\bar h-\bar \delta}^{\bar h+\bar\delta}\text{d}h'\ \text{d}\bar h'\ f^{2}_{\mathcal{O}\mathcal{O}O'}\,.
\end{aligned}
\end{equation}

The set up of this problem is analogous to the one presented in \cite{Pal:2019zzr}. We introduce $\phi_{\pm}$ which bound the indicator function of the rectangular region on $(h,\bar h)$ plane from above and below. For simplicity, we will proceed with the upper bound (the argument for the lower bound is similar) 

\begin{equation}
\mathcal{A} \leq e^{\frac{\beta_L}{2}(h-\delta)+\frac{\beta_R}{2}(\bar h-\bar \delta)}\int dh'\ d\bar h'\ f^{2}_{\mathcal{O}\mathcal{O}O'}\ \phi_{+}(h',\bar h')e^{-\frac{\beta_L}{2}h'-\frac{\beta_R}{2}\bar h'}
\end{equation}

We use the Fourier transform of $\phi_{+}(h',\bar h')$ its bandlimited nature to recast the above in following form
\begin{equation}\label{aux}
\mathcal{A} \leq e^{\frac{\beta_L}{2}(h-\delta)+\frac{\beta_R}{2}(\bar h-\bar \delta)}\int_{-\Lambda}^{\Lambda} dt\ \int_{-\Lambda}^{\Lambda} d\bar t\ e^{-(\beta_L+2\imath t+\beta_R+2\imath\bar t)C/24}p_{\text{aux}}(\beta_L+2\imath t,\beta_R+2\imath \bar t)\ \widehat{\phi}_{+}(t',\bar t')
\end{equation}

 We note that $p_{\text{aux}}(\beta_L+2\imath t,\beta_R+2\imath \bar t)$ doesn't immediately have any modular property. Nonetheless we can make some progress since we are only interested in a asymptotic limit. First of all, note that in \eqref{aux}, $\beta_{L,R}$ and $h,\bar h$, are free parameters. We use this freedom to make the following choice (which we will motivate later) 
 \begin{equation}\label{saddle}
 \beta_{L}=\pi\sqrt{\frac{c-1}{6h}}\,,\ \ \beta_{R}=\pi\sqrt{\frac{c-1}{6\bar h}}\,.
 \end{equation}
 Subsequently, we let $\beta_{L,R}\to 0$ and $h,\bar h\to\infty$ and we want to estimate \eqref{aux} in this limit. To proceed, we need to make some assumptions about the behaviour of $H(h,q)$, that will allow us to replace $p_{\text{aux}}$ with $p(\beta_L,\beta_R)$.  \\
 
\textit{The technical assumption that we make is following:
\begin{equation}\label{assumption}
H(h \to\infty, q_{L} \to 1)= H + \text{error terms} \,,\  H>0\,\ \& \ q_{L}=e^{-\pi\sqrt{\frac{(c-1)}{6h}}}
\end{equation}
and similar statement applies for the right moving part. The value of $H$ depends on the parameter $c$, $h_{\mathcal{O}}$. We shall provide numerical justification for this in \S~\ref{Sec:vir}}. We elaborate further on this assumption in appendix \S \ref{app:assum}.\\

We can now replace\footnote{Technically speaking, we have also assumed that the error term does not accumulate upon summing over $h',\bar h'$.} $p_{\text{aux}}$ in \eqref{aux} with $\frac{1}{H^2}p(\beta_{L}+2\imath t,\beta_R+2\imath \bar t)$ and 
\begin{equation}\label{main}
\mathcal{A} \leq \frac{1}{H^2}e^{\frac{\beta_L}{2}(h-\delta)+\frac{\beta_R}{2}(\bar h-\bar\delta)}\int_{-\Lambda}^{\Lambda} dt\ \int_{-\bar\Lambda}^{\bar\Lambda} d\bar t\ e^{-(\beta_L+2\imath t+\beta_R+2\imath\bar t)C/24}p(\beta_L+2\imath t,\beta_R+2\imath \bar t)\ \widehat{\phi}_{+}(t',\bar t')
\end{equation}

We restrict to the scenario where $h=\bar{h}+J$ and we are approaching $h,\bar h\to\infty$, thus we have $\beta_{L,R}=\pi\sqrt{\frac{C}{6h}}$, note asymptotically $h\simeq \bar h$, hence $\beta_{L}\simeq \beta_R$. Going to the dual channel, we seperate the contribution into the light and the heavy part in the following way (defining $\beta'=\frac{4\pi^2}{\beta+2\imath t}$)
\begin{equation}
\begin{aligned}
p_{L}(\beta'_L,\beta'_R)&=\sum_{h',\bar h'<C/24}f^{2}_{\mathcal{O}\mathcal{O}O'}e^{-\frac{\beta'_L}{2}\left(h'-\frac{C}{24}\right)-\frac{\beta'_R}{2}\left(\bar h'-\frac{C}{24}\right)}H(h',q'_L)H(\bar{h}',q'_R)\\
p_{H}(\beta'_{L},\beta'_R)&=\sum_{h'>C/24\ \text{or}\ \bar h'>C/24}f^{2}_{\mathcal{O}\mathcal{O}O'}e^{-\frac{\beta'_L}{2}\left(h'-\frac{C}{24}\right)-\frac{\beta'_R}{2}\left(\bar h'-\frac{C}{24}\right)}H(h',q'_L)H(\bar{h}',q'_R)
\end{aligned}
\end{equation}
and analyze them sperately.

$\bullet$ Light part: We have to evaluate the integral of the following from 
\begin{equation}
\begin{aligned}
&\int_{-\Lambda}^{\Lambda} dt\ \int_{-\bar\Lambda}^{\bar\Lambda} d\bar t\ \left(\frac{4\pi^2}{(\beta_L+2\imath t)(\beta_R+2\imath t)}\right)^{\nu/2+1/4}\times \\
&\ \exp\left[\frac{\pi^2(c-1)}{12(\beta_L+2\imath t)}+\frac{\pi^2(c-1)}{12(\beta_R+2\imath t)}+2\imath(t+\bar t)(c-1)/24\right] \widehat{\phi}_{+}(t',\bar t')
\end{aligned}
\end{equation}
This integral can be evaluated by saddle point method with saddle being $t=\bar t=0$ and gives
$$4\pi^2\widehat{\phi}_{+}(0)\left(\frac{4\pi^2}{\beta_L\beta_R}\right)^{\nu/2+1/4}\frac{3\beta_L^{3/2}\beta_R^{3/2}}{8\pi^3(c-1)}\exp\left[\frac{\pi^2(c-1)}{12\beta_L}+\frac{\pi^2(c-1)}{12\beta_R}\right]$$

Using $\beta_L=\beta_R=\beta$, contribution to \eqref{main} from the light part comes out to be
\begin{equation}\label{lighthh}
4\pi^2\widehat{\phi}_{+}(0)\frac{1}{H^2}\left(\frac{2\pi}{\beta}\right)^{\nu+1/2}\frac{3\beta^3}{8\pi^3(c-1)}e^{\beta h+\frac{\pi^2(c-1)}{6\beta}}
\end{equation}

$\bullet$ Heavy part: to estimate the heavy part, we first bound it by its absolute value
\begin{equation}
\begin{aligned}
I&\equiv \int_{-\Lambda}^{\Lambda} dt\ \int_{-\bar\Lambda}^{\bar\Lambda} d\bar t\ \\
 &\left(\frac{2\pi}{\sqrt{\beta^2+4t^2}}\right)^{\nu+1/4}\sum' f^{2}_{\mathcal{O}\mathcal{O}O'}e^{-\frac{2\pi^2\beta}{\beta^2+4t^2}\left(h'-\frac{C}{24}\right)-\frac{2\pi^2\beta}{\beta^2+4\bar t^2}\left(\bar h'-\frac{C}{24}\right)}H(h',q'_L)H(\bar{h}',q'_R) |\widehat{\phi}_{+}(t',\bar t')|
 \end{aligned}
\end{equation}
and then use the fact that one of $h$ or $\bar{h}$ is bigger than $C/24$. The prime over the sum reminds us that we are dealing with the heavy sector. Thus (without loss of generality assuming, $h'<C/24$) 
\begin{equation}
\begin{aligned}
I&\leq e^{\frac{2\pi^2}{\beta}C/24}\times \\
&\int_{t,\bar t}\left(\frac{4\pi^2}{\sqrt{|\beta_L(t)\beta_R(t)|}}\right)^{\nu/4+1/2}\sum' f^{2}_{\mathcal{O}\mathcal{O}O'}e^{-\frac{2\pi^2\beta}{\beta^2+4t^2}\left(h'\right)-\frac{2\pi^2\beta}{\beta^2+4\bar t^2}\left(\bar h'-\frac{C}{24}\right)}H(h',q'_L)H(\bar{h}',q'_R) |\widehat{\phi}_{+}(t',\bar t')|
\end{aligned}
\end{equation}
Now the integral in the right hand side is dominated by $t\sim\bar t\sim O(1)$ number, since  increasing $t$ maximizes the integrand in exponential manner. As a result in the $\beta\to 0$ limit, the effective temperature becomes large and we approximate it by the dual channel\footnote{We assume that the error due to this approximation does not accumulate, since the integral is taken over a finite region, we are safe to make such assumptions. Similar assumption is also present in the fixed spin analysis in \cite{Mukhametzhanov:2020swe}.}. Thus we have
\begin{equation}
\begin{aligned}
&I\leq O\left( \beta^{-\nu-1/2}e^{\frac{2\pi^2}{\beta}c/24}\int_{t,\bar t}p\left(\frac{4t^2}{\beta},\frac{4\bar t^2}{\beta}\right)|(\Lambda-t)(\bar\Lambda-\bar t)|\right)\\
&=O\left(\beta^{-\nu-1/2}\int_{t,\bar t}e^{\frac{(\pi^2+t^2+\bar t^2)C}{12\beta}}|(\Lambda-t)(\bar\Lambda-\bar t)|\right)\\
&\simeq O\left(\beta^{-\nu-1/2+4}e^{\frac{(\pi^2+\Lambda^2+\bar\Lambda^2)C}{12\beta}}\right)
\end{aligned}
\end{equation}

Thus the contribution of the heavy part to \eqref{main} is given by
\begin{equation}\label{heavyhh}
O\left(\beta^{-\nu-1/2+4}e^{\frac{(\pi^2+\Lambda^2+\bar\Lambda^2)C}{12\beta}+\beta h}\right)
\end{equation}
Comparing \eqref{lighthh} and \eqref{heavyhh} we have suppresion if we choose
\begin{equation}\label{cons}
\Lambda^2+\bar\Lambda^2=\pi^2
\end{equation}
For a square like support, we can choose $\Lambda=\bar\Lambda=\frac{\pi}{\sqrt{2}}$. Similar considerations would follow for the lower bound as well. Combining everything and optimizing $\beta$ as a function of $\Delta$, we have
\begin{equation}
\begin{aligned}\label{primary}
&4\pi^2\widehat{\phi}_-(0)\left[\frac{1}{4H^2}  \left(\frac{c-1}{24}\right)^{1/4-\nu/2} (h\bar h)^{\nu/4-5/8}e^{\pi\sqrt{\frac{(c-1)h}{6}}+\pi\sqrt{\frac{(c-1)\bar h}{6}}}\right]\\
&\leq \mathcal{A}\leq\\
&4\pi^2\widehat{\phi}_+(0) \left[\frac{1}{4H^2}  \left(\frac{c-1}{24}\right)^{1/4-\nu/2} (h\bar h)^{\nu/4-5/8}e^{\pi\sqrt{\frac{(c-1)h}{6}}+\pi\sqrt{\frac{(c-1)\bar h}{6}}}\right]
\end{aligned}
\end{equation}

Modulo the order one number, the above result matches with that appearing in \cite{Collier:2018exn,Collier:2019weq}. 

\subsection{Asymptotic spectral gap in OPE channel} 
The notion of spectral gap for the analysis on $(h,\bar h)$ plane requires more sophistication. Let us go back and restate the notion of spectral gap in $\Delta$ variable. Imagine considering intervals of length $2\delta$, centered at $\Delta$. The asymptotic spectral gap corresponds to minimum value of $\delta$ such that the interval contains some states given any CFT. On $(h,\bar h)$ plane, we can instead consider $2$-dimensional objects centred around a point. Then the question is how much can we shrink the object so that it still contains some operators as we move centre point out to infinity. For example, one can consider circles or ellipses and send the center to infinity, a measure of spectral gap would be the radius $r$ of the circle or semi-minor length $a$ of the ellipse. Now each choice of $\phi_-$ will provide us with some measure of  asymptotic spectral gap (or an lower bound on $r$ and/or $a$), the information about the shape of the geometric object is contained in $\phi_-$. 

The problem just described is linked with judicious choice of $\phi_-$ such that
\begin{equation}
\phi_- \leq \Theta_{\text{circle/ellipse}}
\end{equation}
where $\Theta_{\text{circle/ellipse}}$ is the indicator function of circle/ellipse i.e. it takes value $1$ for points within the circle/ellipse and $0$ otherwise. Here, $\widehat{\phi}_-$ has bounded support satisfying \eqref{cons}.  The question is what is the minimum possible value of the radius of the circle or semi-minor length of the ellipse such that $\widehat{\phi}\geq 0$. This minimum possible value would be an upper bound on the minimal value of $r$ or $a$.\\

\begin{figure}[!ht]
\centering
\includegraphics[scale=0.65]{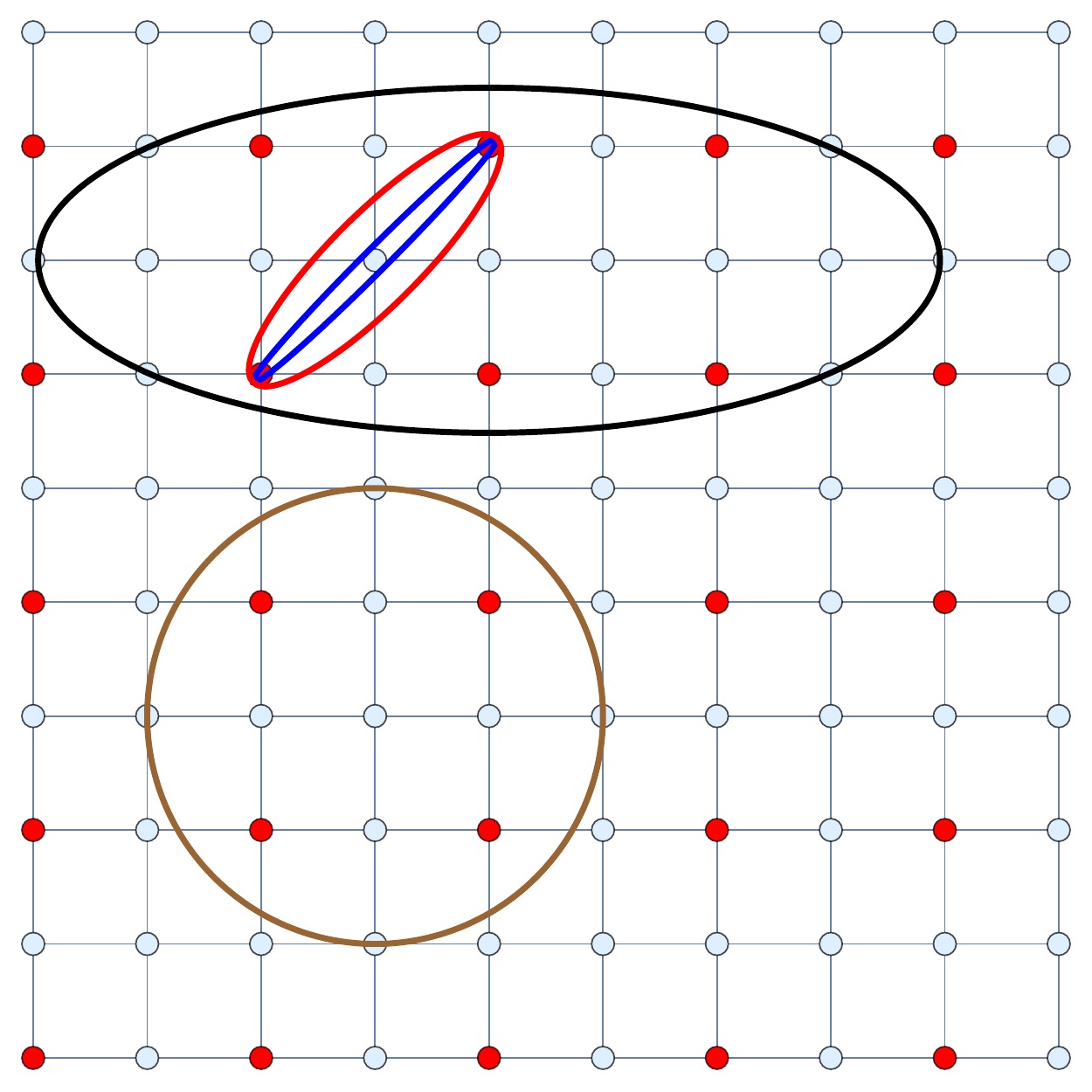}
\caption{The figure depicts the $(h',\bar h')$ plane. The lines parallel to $h'=\bar h'$ line are fixed spin lines. The lines perpendicular to fixed spin lines are fixed $\Delta$ lines. Suppose, we find an example such that all the operators appearing in the OPE of two $\mathcal{O}$ operators are separated by $\delta h=\delta\bar{h}=2$. The red dots in the figure depcits them. The Tauberian analysis tells us that the brown circle ($r=2$) and the black ellipse ($\text{semi-minor  axis}>\sqrt{2}$) will always have some operators within it, as the center approaches infinity. The black ellipse can be rotated by $\pi/4$, the same result holds. To obtain the red and the blue ellipse, one has to take the black ellipse with semiminor axis of length $\sqrt{2}$ and squeeze the other axis, this can only be achived by using $T$ invariance in the fixed spin analysis. In the limit, the blue ellipse degenerates into a line segment of length (defined as Euclidean distance on the $(h',\bar h')$ plane) $2\sqrt{2}$ (or in terms of $\Delta$, the length of the segment is $\delta\Delta=4$). This is the bound on the gap that we obtain from fixed spin Tauberian analysis and this also proves the optimality of our bound, since we can not push the bound to a smaller value. An explicit example of such kind involves 3 copies of Ising, analyzed with respect to the diagonal Virasoro.}
\label{figure1}
\end{figure}

We can arrive at three related results (\ref{figure1}) using the above chain of logic, followed by aprropriate choice of $\phi_-$:\\

\begin{enumerate}
\item $\label{1}\bigstar$ \textit{Given any circular region of radius $r>2$, centered at $(h,\bar h)$, as the center approahes $\infty$, the circle always contains states appearing in the intermediate channel upon fusing $\mathcal{O}$ with $\mathcal{O}$. This implies $r_{optimal}\leq 2$.}\\

\item $\label{2}\bigstar$  \textit{Given any elliptical region with semiminor axis of length $a>\sqrt{2}$, parallel to $h'=0$ or $\bar{h}'=0$ line, centered at $(h,\bar h)$; as the center approahes $\infty$, the ellipse always contains states appearing in the intermediate channel upon fusing $\mathcal{O}$ with $\mathcal{O}$. This implies $a_{optimal}\leq \sqrt{2}$.}\\

\item $\label{3}\bigstar$  \textit{Given any elliptical region with semiminor axis of length $a>\sqrt{2}$, parallel to fixed spin line or its perpendicular, centered at $(h,\bar h)$; as the center approahes $\infty$, the ellipse always contains states appearing in the intermediate channel upon fusing $\mathcal{O}$ with $\mathcal{O}$. This implies $a_{optimal}\leq \sqrt{2}$}\\
\end{enumerate}

It turns out that the above bounds are not the optimal/minimal ones, since one can construct the following situation where $r<2$ and $a<\sqrt{2}$, and still the circle/ellipse contains operators. Consider $3$ copies of Ising model and decompose the operator content in terms of left diagonal Virasoro and right diagonal Virasoro (rather than 3 copies of Virasoro, which is bigger algebra). Since $c_{eff}>1$, we have infinite number of primaries with respect to the diagonal Virasoro. Then one can consider $4$ point correlator of $\epsilon\otimes\epsilon\otimes\epsilon$, viewing it as a primary operator with respect to the diagonal Virasoro. The primary operators appearing in the OPE has a gap of $\delta h=\delta \bar h=2$ (the primaries are of the form $p_1\otimes p_2\otimes p_3$ where at least one of $p_i$ is $\mathbb{I}$ and rest of the two has even  $h$ and even $\bar h$, note that they are primaries with respect to diagonal Virasoro, not with respect to the 3 copies of Virasoro.) In this case both the holomorphic as well as the anti-holomorphic conformal dimensions are gapped by even integers. This immediately imply $a_{\text{optimal}}\leq 1 <\sqrt{2}$ and $r_{\text{optimal}}\leq \sqrt{2}\leq 2$. 
This shows that the bound we have achieved is not near the optimality. Later we will show by using $T$ invariance and doing the fixed spin analysis, that the bound on gap is indeed $\delta h=\delta \bar h=2$ (or $\delta\Delta=4$ for fixed spin). Before delving into that let us show how to arrive at $\bigstar\ 1, \bigstar\ 2,\bigstar\ 3$.\\

To obtain $\bigstar 1$, we choose $\phi_-(h',\bar h')=f(h'-h,\bar h-\bar h') $, \cite{berdysheva1999two,sb} where, 
\begin{equation}\label{f}
f(x,y)=\frac{1}{2}\left(2-\frac{x^2+y^2}{2}\right)\left(\frac{\cos\left(\frac{\pi x}{2\sqrt{2}}\right)\cos\left(\frac{\pi y}{2\sqrt{2}}\right)}{\left(1-\frac{x^2}{2}\right)\left(1-\frac{y^2}{2}\right)}\right)^2
\end{equation}
The function $\phi_-(h',\bar h')$ bounds a circular region centered at $(h,\bar h)$ with a radius of $2$ on $(h',\bar h')$ plane. The support of the Fourier transform of the function can be deduced using Paley-Weiner theorem of multivariable complex analysis and comes out to be $\frac{\pi}{\sqrt{2}}$.\\

To obtain $\bigstar 2$, we generalize the above analysis by allowing $\Lambda\neq \bar\Lambda$. One can achieve this by scaling the $x$ and $y$ variable asymmetrically in \eqref{f}, the resulting function will bound the indicator function of an ellipse, rather a circle. We still have to satisfy 
$$\Lambda^2+\bar\Lambda^2=\pi^2\,.$$
We use the following function
\begin{equation}\label{f2}
f_{\ell}(x,y)=\frac{1}{2}\left(2-\frac{\Lambda^2 x^2+\bar\Lambda^2 y^2}{\pi^2}\right)\left(\frac{\cos\left(\frac{\Lambda x}{2}\right)\cos\left(\frac{\Lambda y}{2}\right)}{\left(1-\frac{\Lambda^2x^2}{\pi^2}\right)\left(1-\frac{\bar\Lambda^2 y^2}{\pi^2}\right)}\right)^2
\end{equation}
The $f_{\ell}$ bound $\Theta(\Lambda^2 x^2+\bar\Lambda^2 y^2<2\pi^2)$
, an ellipse with semimajor and semi minor axis given by $a=\frac{\sqrt{2}\pi}{\Lambda}$ and $b=\frac{\sqrt{2}\pi}{\bar \Lambda}$, the constraint on $\Lambda$ and $\bar\Lambda$ translates into
$$a^{-2}+b^{-2}=1/2$$. 

To arrive at $\bigstar 3$, we further generalize to a rotated version ($\pi/4$ rotation) of the previous ellipse. This can be achieved by working with variables
\begin{equation}
\begin{aligned}
\widetilde{t}=\frac{t+\bar{t}}{\sqrt{2}}\,\ &\&\  \widetilde{\bar t}=\frac{t-\bar t}{\sqrt{2}}\,,\\
\widetilde{x}=\frac{x+y}{\sqrt{2}}\,\ &\&\ \widetilde{y}=\frac{x-y}{\sqrt{2}}
\end{aligned}
\end{equation}
and now we have 
\begin{equation}
\widetilde{\Lambda}^2+\widetilde{\bar\Lambda}^2=\pi^2
\end{equation}
This will bound an ellipse whose one axis is parallel to fixed spin lines i.e $h-\bar h=\text{constant}$. Again the length of semimajor and semi minor axis will satisfy $a^{-2}+b^{-2}=1/2$.\\

\subsection{Fixed spin analysis}
Fixed spin analysis is done by projecting the $p_{\text{aux}}$ onto fixed spin sector by another finite Fourier integeral. We recall that only even spins appear in the expression for $p_{\text{aux}}(\beta,\omega)$. So we introduce a new variable 
$$j\equiv \frac{1}{2}|h-\bar h|$$ 
The spin projected partition function can be written as
\begin{equation}
\begin{aligned}
p_{j}(\beta)&=\frac{1}{1+\delta_{j,0}}\int_{-1/2}^{1/2} d\omega\ \left(e^{2\pi\imath\omega j}+e^{-2\pi\imath\omega j}\right) p_{\text{aux}}(\beta,\omega)\\
&=\int d\Delta\ a(\Delta,j)e^{-\frac{\beta}{2}(\Delta-C/12)}
\end{aligned} 
\end{equation}
Now again we replace $p_{\text{aux}}(\beta,\omega)$ with the actual $p$ and assume that the error term in $H$ does not accumulate under this projection. Rest of the calculation proceeds in exaclty same way as in \cite{Mukhametzhanov:2020swe}. We only need to rescale $\beta\to\beta/2$, which provides us with $\Lambda\leq \pi/2$. The zero modes of Selberg-Beurling function (for a detailed expositions to these functions, see \cite{Mukhametzhanov:2020swe})  gives
\begin{equation}
2\pi\widehat{\phi}_{\pm}(0)=2\delta\pm \frac{2\pi}{\Lambda}
\end{equation}
Choosing $\Lambda=\frac{\pi}{2}$, we have
\begin{equation}
\begin{aligned}
&\frac{2(\delta-2)}{1+\delta_{j,0}}\left[\frac{1}{4H^2}  \left(\frac{c-1}{24}\right)^{1/4-\nu/2} \left(\frac{\Delta}{2}\right)^{\nu/2-5/4}e^{\pi\sqrt{\frac{(c-1)\Delta}{3}}}\right] \\
&\leq \mathcal{A}_{j}\leq\\
& \frac{2(\delta+2)}{1+\delta_{j,0}}\left[\frac{1}{4H^2}  \left(\frac{c-1}{24}\right)^{1/4-\nu/2} \left(\frac{\Delta}{2}\right)^{\nu/2-5/4}e^{\pi\sqrt{\frac{(c-1)\Delta}{3}}}\right]
\end{aligned}
\end{equation}

We can arrive at the optimal asymptotic spectral gap using the lower limit:\\

$\bigstar \ 4$: \textit{Given a fixed spin line, the optimal asymptotic spectral gap between consecutive primaries in $\Delta$ variable is $4$.}\\

This precisely comes from the conjectured scenario we described before, see the degenerate blue ellipse in the figure~\ref{figure1}, this would in the limit become a line segment of length $4$ (i.e $\delta\Delta=4$) 

$\bullet$ There is another slick way to see this result. Basically, while doing the fixed spin analysis, we are effectively setting one of axis length of the ellipse to be $0$. In the $h',\bar h'$ analysis, we can not do that, since this would violate $a^{-2}+b^{-2}\leq 1/2$ bound. Using the extra information that spins are integers (in this case even integers) we can let $b\to0$ and $a\to \sqrt{2}$. This would correspond to the degenerate blue ellipse in the figure~\ref{figure1}.

All the geometrical idea presented in this section naturally goes over to the discussion of Cardy density of states on the plane and will improve some of discussion related to optimal gap in \cite{Pal:2019zzr}. We will report that in detail somewhere else.

\section{Asymptotics and Numerics of Virasoro Block}
\label{Sec:vir}
The aim of this section to study the asymptotics of the $q$ expansion coefficient of $H(h,q)$ defined in \eqref{eq:Vh} in order to substantiate the assumptions used to get the Tauberian bounds. We begin with 
\begin{equation}
H(h,q)=1+\sum_{n}c_n(h) q^{n}
\end{equation}
By Zamolodchikov recurssion relation, it is possible to evaluate $c_n$ numerically. Let us briefly review it.

One efficient approach to evaluate $H(h,q)$ (and hence the Virasoro conformal block) is the {\it Zamonodchikov recursion relation}. This recursion relation is first derived in  \cite{zamu, Zamolodchikov1984}, developed by  \cite{Cho2017a} and recently used in the context of the bootstrap \cite{BaeLeeLee2016,CollierKravchukLinYin2017,EsterlisFitzpatrickRamirez2016,LinShaoSimmons-DuffinWangYin2017} and the calculation of the entanglement entropy or the OTOC \cite{Kusuki:2019gjs, Ruggiero2018, Chang2018}. We will briefly review it.

Let us consider the following series expansion,
\be\label{eq:seriesH}
H^{21}_{34}(h_p|q)=1+\sum_{k=1}^\infty c_k(h_p) q^{k},
\ee
where $H^{21}_{34}$ is the generalization of the funsction $H$ in (\ref{eq:Vh}) with external conformal dimensions $h_i$ ($i=1,2,3,4$).
The coefficients $ c_k(h_p)$ can be calculated by the following recursion relation,
\begin{equation}\label{eq:ck}
	c_k(h_p) = \sum_{i=1}^k \sum_{\substack{m=1, n=1\\mn=i}} \frac{R_{m,n}}{h_p - h_{m,n}} c_{k-i}(h_{m,n}+mn),
\end{equation}
where $R_{m,n}$ is a constant in $h_p$, which is defined by
{\scriptsize
\begin{equation}\label{eq:Rmn}
R_{m,n}=2\fr{
\substack{m-1\\ \displaystyle{\prod} \\p=-m+1\\p+m=1 (\text{mod } 2) \  } \ 
\substack{n-1\\ \displaystyle{\prod} \\q=-n+1\\q+n=1 (\text{mod } 2) }
\pa{\la_2+\la_1-\la_{p,q}}\pa{\la_2-\la_1-\la_{p,q}}\pa{\la_3+\la_4-\la_{p,q}}\pa{\la_3-\la_4-\la_{p,q}}}
{\substack{
\substack{m \\ \displaystyle{\prod} \\k=-m+1 } \ \ 
\substack{n \\ \displaystyle{\prod} \\l=-n+1 }\\
(k,l)\neq(0,0), (m,n)
}
 \lambda_{k,l}}.
\end{equation}
}
In the above expressions, we used the notations,
\begin{equation}
\begin{aligned}
&c=1+6\pa{b+\fr{1}{b}}^2,  \hspace{16ex}   h_i=\fr{c-1}{24}-\la_i^2,\\
&h_{m,n}=\fr{1}{4}\pa{b+\fr{1}{b}}^2-\lambda_{m,n}^2,  \hspace{10ex}  \lambda_{m,n}=\fr{1}{2} \pa{\fr{m}{b}+nb}.
\end{aligned}
\end{equation}

We wil mostly restrict to the case where all the operators are same and investigate the behavior of $c_n$ as $n\to \infty$. Even though this seems to fall naturally under the umbrella of Tauberian formalism, the positivity of the coefficients $c_n$ is not guaranteed to begin with. Thus we numerically probe the behavior of $c_n$ for various values of central charge.

\subsection{Positivity/sign indefiniteness of $c_n$ from numerics}
First of all, we can comment that the positivity of $c_n$ can be verified in the large $c$ limit from the following equation \cite{Kusuki2018},
	 \begin{equation}
		c_{2m}\ar{c \to \infty}\fr{1}{m!} \BR{\fr{c}{2} \pa{1-\fr{16}{c}\Delta_\O}}^2m \ \ \ \ \text{for } 2m\ll c.
	\end{equation}
Therefore, we will focus on the small $c$ regime.
Note that more generally, if we consider a correlator $\vev{  \O_A(0)\O_A(z,\bar{z} )\O(1)_B\O_B(\infty) }$ with conformal dimensions $h_A=\bar{h}_A$ and $h_B=\bar{h}_B$, we have
	 \begin{equation}\label{eq:cnlim}
		c_{2m}\ar{c \to \infty}\fr{1}{m!} \BR{\fr{c}{2} \pa{1-\fr{32}{c}h_A}\pa{1-\fr{32}{c}h_B}}^m \ \ \ \ \text{for } 2m\ll c.
	\end{equation}

\begin{figure}[t]
\centering
  \includegraphics[width=7.0cm,clip]{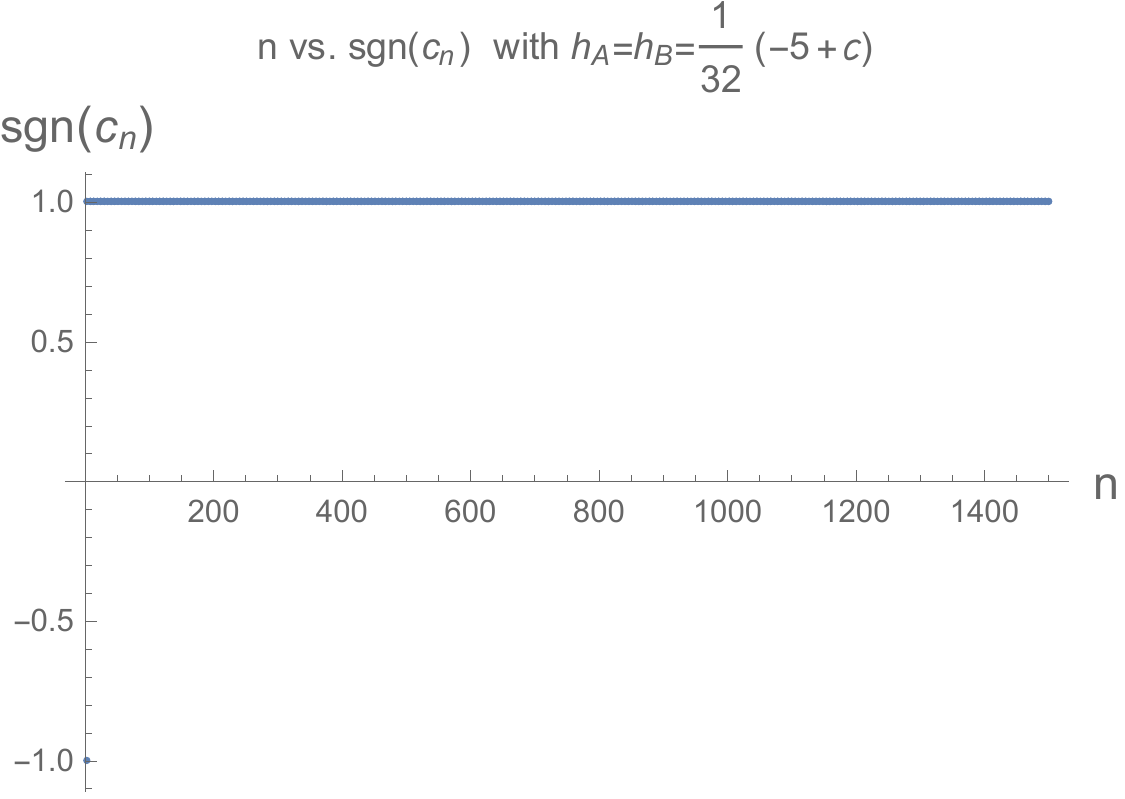}
  \includegraphics[width=7.0cm,clip]{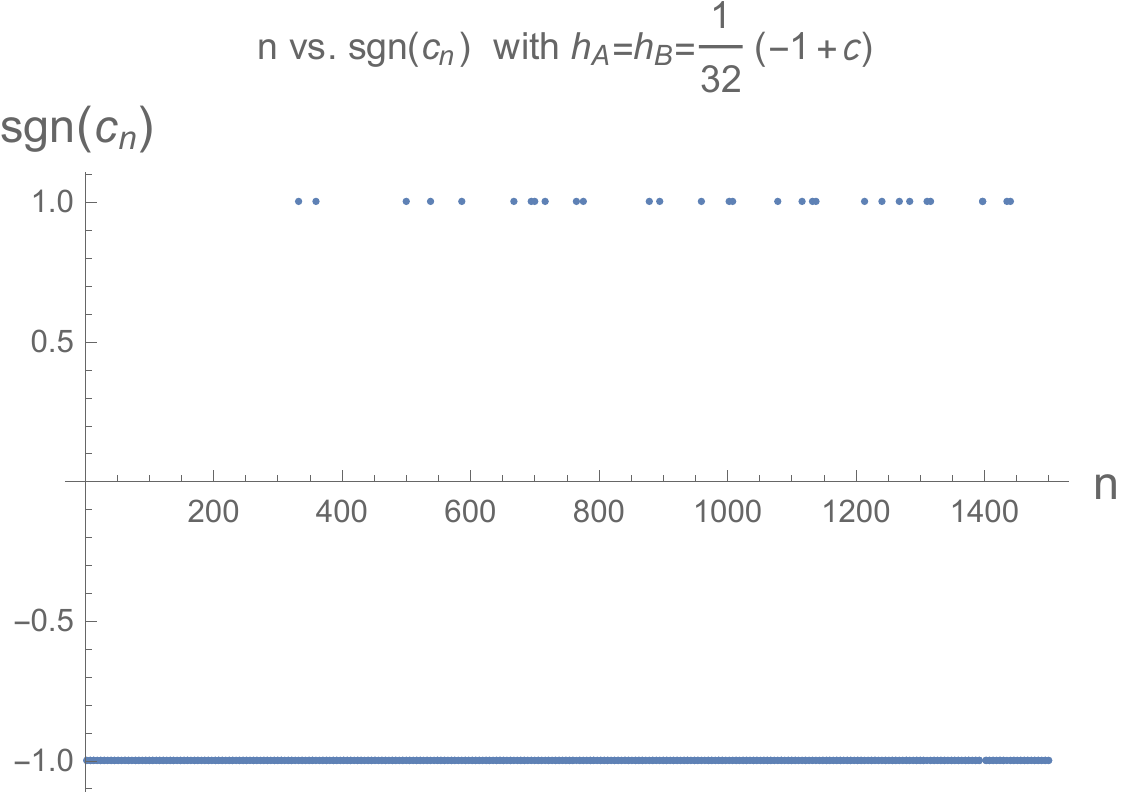}
  \includegraphics[width=7.0cm,clip]{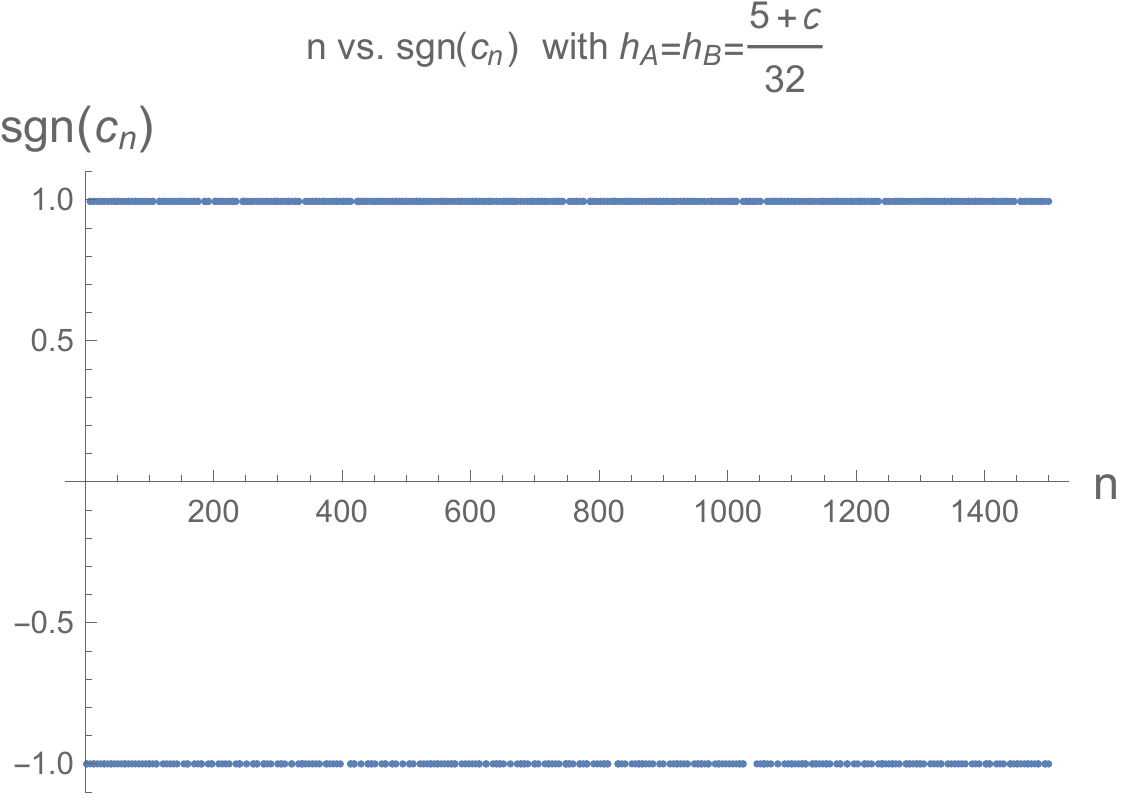}
  \includegraphics[width=7.0cm,clip]{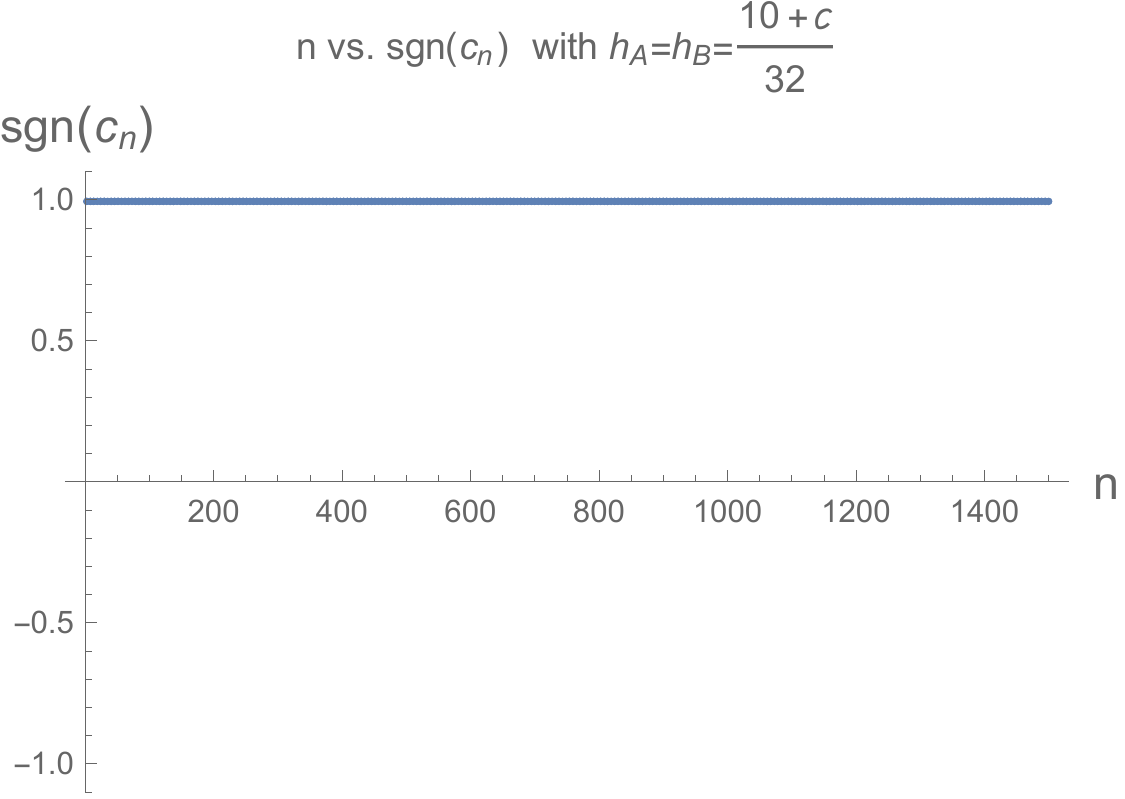}
 \caption{The $n$-dependence of $\sgn(c_n)$ for various $h_\O=\fr{c-5}{32},\fr{c-1}{32},\fr{c+5}{32}$, and $\fr{c+10}{32}$ with $c=30$.
One can find that the $c_n$ away from the region $h_\O \in [ \fr{c-1}{32}, \fr{c+5}{32}  ]$ is always positive, whereas the $c_n$ in the region $h_\O \in [ \fr{c-1}{32}, \fr{c+5}{32}  ]$ can be negative.
}
 \label{fig:ex}
\end{figure}
In Figure \ref{fig:ex}, we show some plots for the $n$-dependence of the coefficients $c_n$ with $c=30$.
We find that in the region $h_\O \in [ \fr{c-1}{32}, \fr{c+5}{32}  ]$ (with $2h_\O=\Delta_\O$), the coefficients $c_n$ can be negative, while the coefficients $c_n$ are always positive away from the region $[ \fr{c-1}{32}, \fr{c+5}{32}  ]$.
It means that without the large $c$ limit (i.e when $c$ is not the largest parameter in the problem), there is a region where $c_n$ can be negative even if $h_A=h_B$.
\begin{figure}[t]
\centering
  \includegraphics[width=7.0cm,clip]{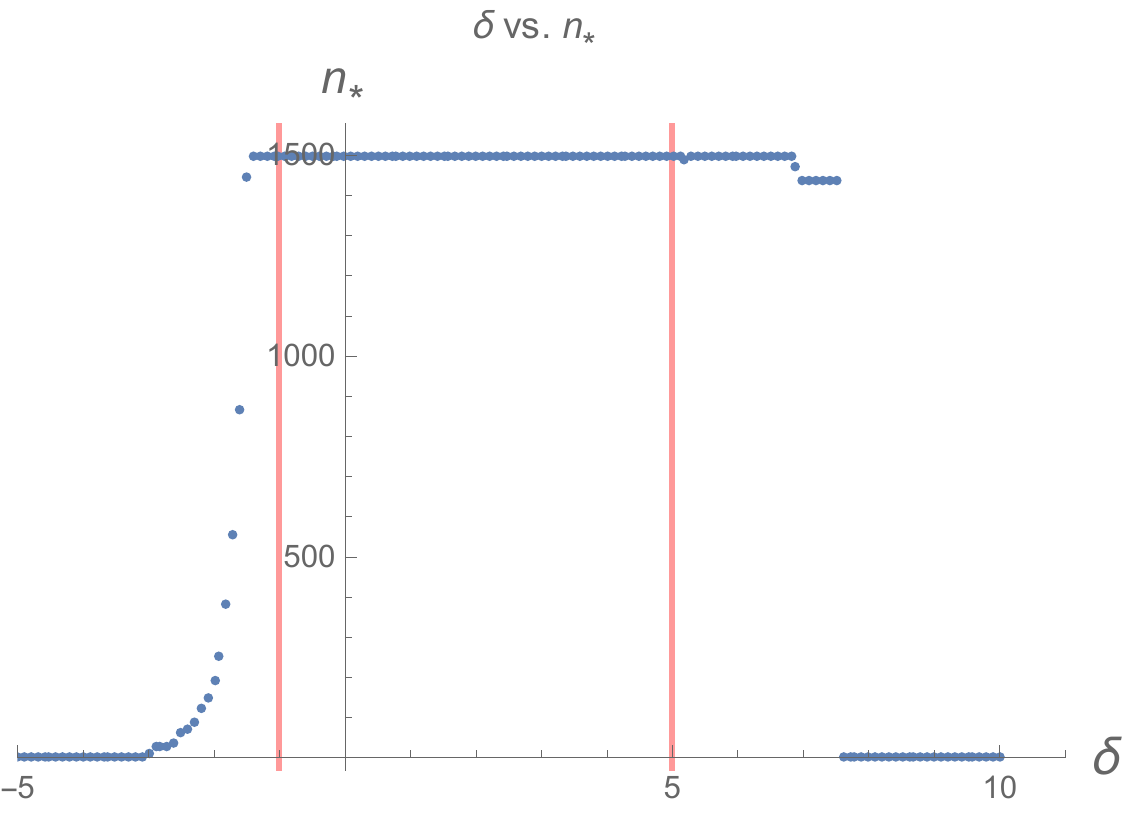}
  \includegraphics[width=7.0cm,clip]{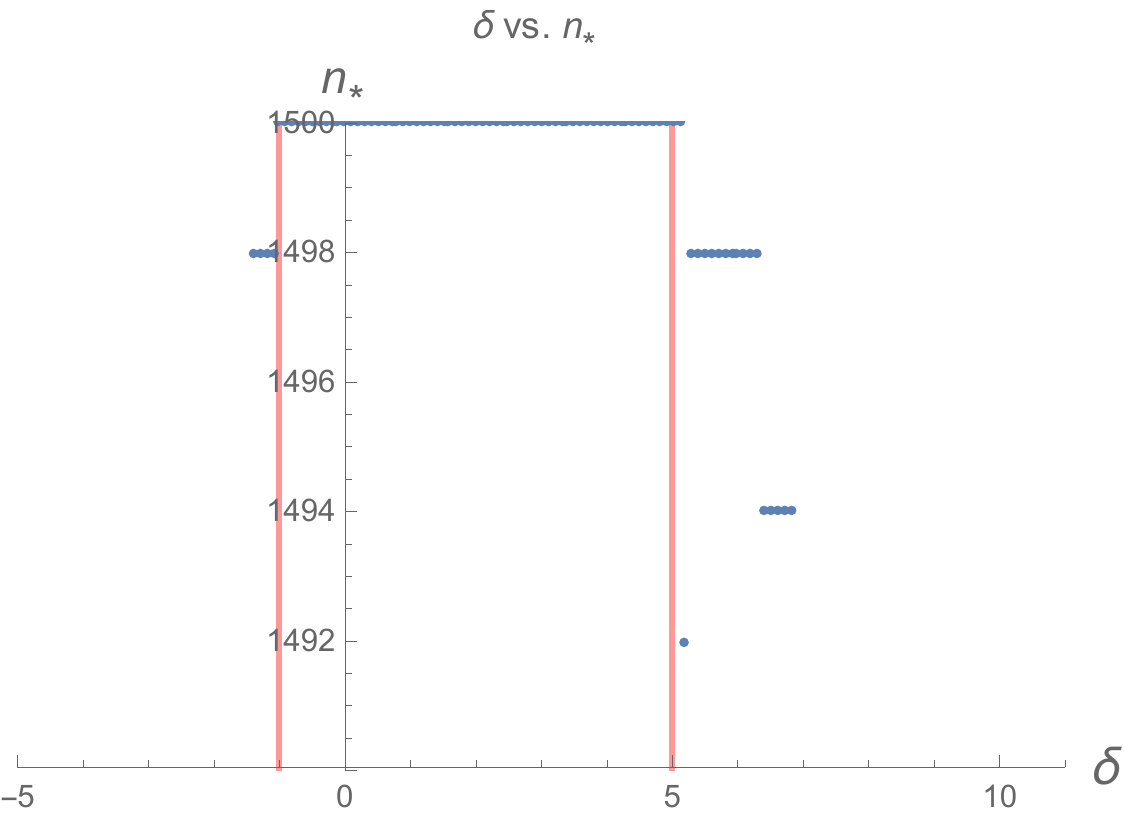}
 \caption{The blue dots show the maximal value $n_* (\leq 1500)$ $s.t.$ $c_{c_*}<0$.
The red lines show the special values $\d=-1,5$.
Here we fix $c=30$ and define $\fr{c+\d}{32}\equiv h$.
The right figure is the zoomed version of the left figure.
One can find the plateau in the region  $\d \in [-1,5]$.
Note that the upper bound $n=1500$ just comes from the limitation of our machine power.
We expect that one can see the plateau in $\d \in [-1,5]$ more clearly if we take higher $n$ terms into account.
}
 \label{fig:first}
\end{figure}
To see more detailed patterns, we search for the maximal $n_*  \ \ s.t. \ \ c_{n_*}<0 \ \ $ up to $n\leq 1500$.
The motivation is that if $n_*$ is very close to $n=1500$, it would imply that the negative coefficients appear many times in the range $(c_2 , c_{1500})$. 
In Figure \ref{fig:first}, we give $n_*$ as a function of the parameter $\d$, defined in terms of conformal dimension of external operator $\O$ i.e.\!~ $h_\O=\fr{c+\d }{32}$. It obviously shows the specialty of the values $h_\O=\fr{c-1}{32}, \fr{c+5}{32}$ (red lines in the figure),
which are the edges of the plateau $[\fr{c-1}{32}, \fr{c+5}{32}]$.
From this observation, we can conjecture,
\begin{itemize}
\item if $h_\O \in [ \fr{c-1}{32}, \fr{c+5}{32}  ]$, $\sgn(c_n)$ oscillates frequently,
\item if $h_\O \notin [ \fr{c-1}{32}, \fr{c+5}{32}  ]$, the negative coefficient does not appear so many times.
\end{itemize}
If the above conjectures are true, then we might expect that a few numbers of the negative coefficients can have a  negligible effect to the evaluation of asymtptotic behaviors of the Virasoro block.
\begin{figure}[t]
\centering
  \includegraphics[width=7.0cm,clip]{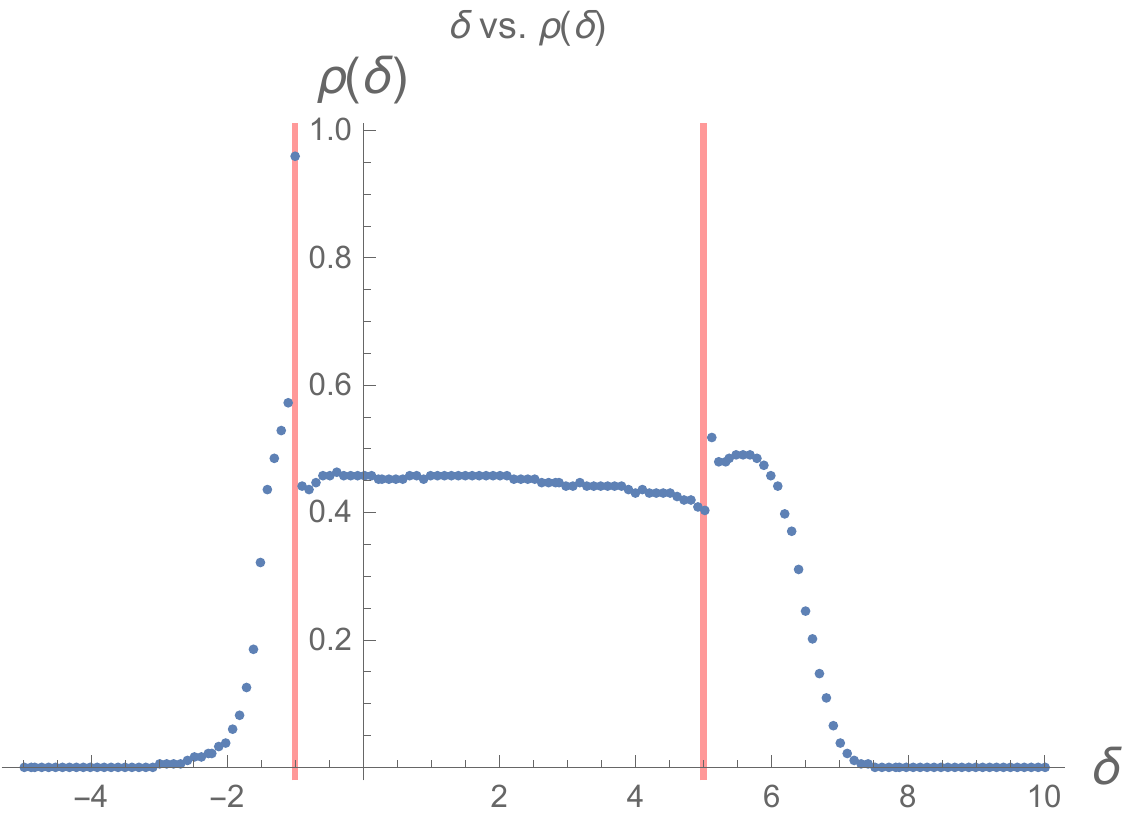}
 \caption{The blue dots show the $\d$-dependence of the density $\rho(\d)$ (see  (\ref{eq:den1})). 
The red lines show the special values $\d=-1,5$.
Here we also fix $c=30$.
}
 \label{fig:density}
\end{figure}
To justify our expectation, we will focus on a {\it density} of the negative coefficients.
We first define the density as
\begin{equation}\label{eq:den1}
\rho(\d) \equiv \fr{\# \text{ of negative coefficients in } [c_2 , c_{1500} ] }{750}.
\end{equation}
We show the $\d$-dependence of the density $\rho(\d)$ in Figure \ref{fig:density}, which implies,
\begin{itemize}
\item
$\rho(\delta) \sim \fr{1}{2}$, if $\d \in  [ -1, 5  ]$. This means that the number of the positive coefficients and the negative coefficients are almost equal to each other.

\item
$\rho(\delta)$ decreases as $\d$ is apart from $ [ -1, 5  ]$.
The strange behavior near $\d=-1,5$ might be due to a lack of precision in $n$.
If we increase the cut off $n$ to higher than $1500$, one would find the plateau clearly.

\item
The value $\d=-1$ is special, which shows $\rho(-1) \sim 1$. This specialty comes from the $h=\fr{c-1}{32}$ transition \cite{Kusuki:2019gjs,kusuki1,Kusuki2018c,Kusuki2018, Brehm:2019pcx} and can be obviously found in Figure \ref{fig:ex}.

 \end{itemize}

\begin{figure}[t]
\centering
  \includegraphics[width=7.0cm,clip]{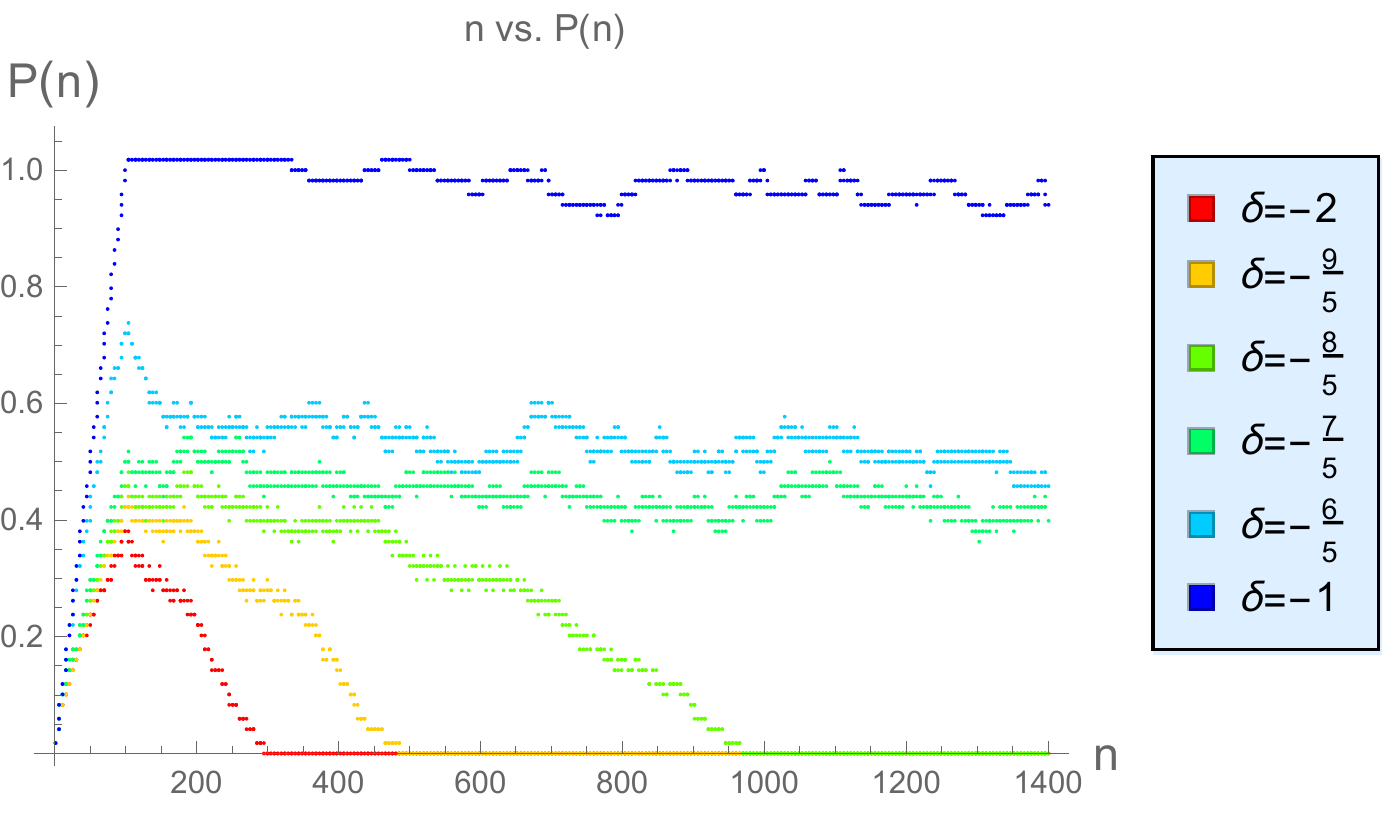}
  \includegraphics[width=7.0cm,clip]{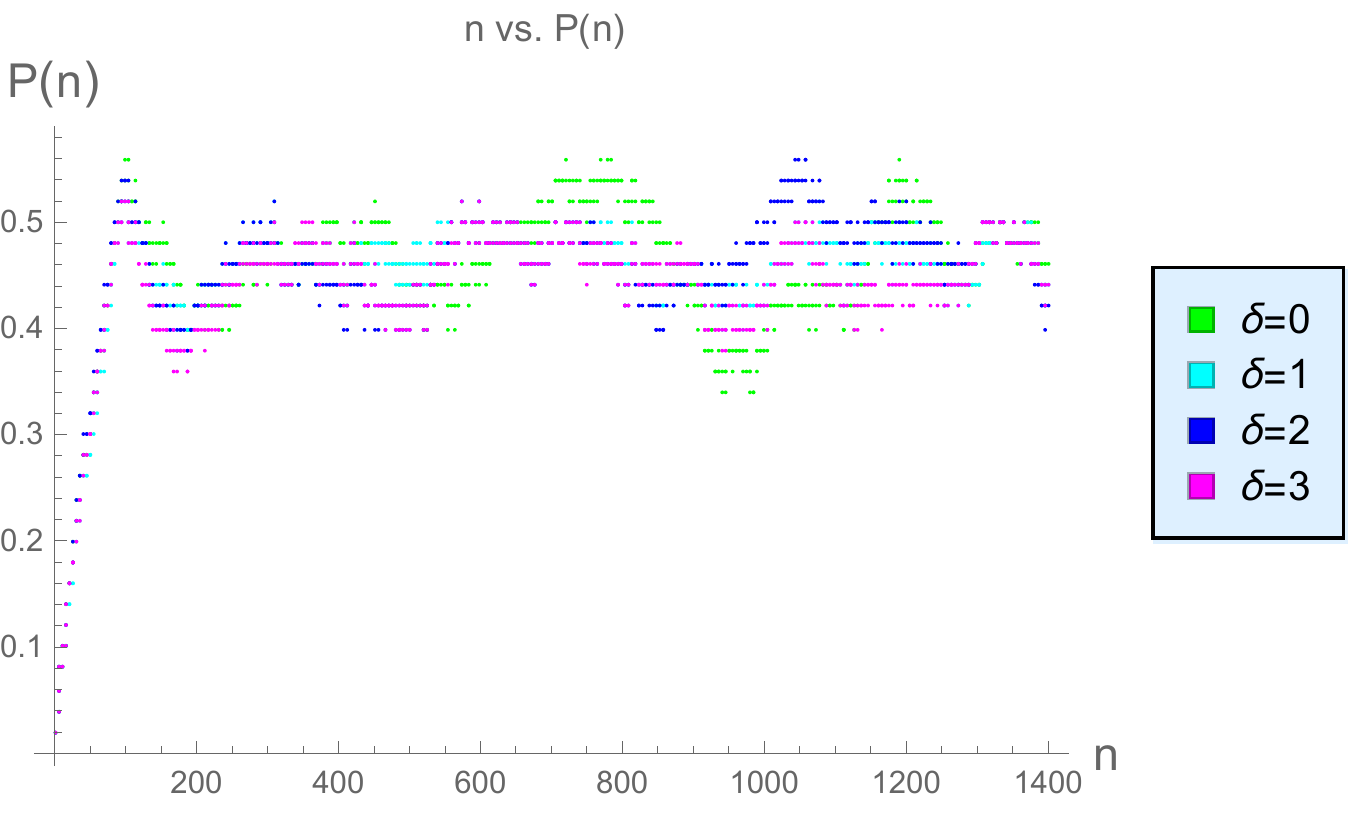}
 \caption{Upper: The $n$-dependence of $P(n)$ for various $\d=-2,-\fr{9}{5}, -\fr{8}{5}, -\fr{7}{5}, -\fr{6}{5}$, and $-1$ with $c=30$ (see (\ref{eq:den2})). 
Lower: The $n$-dependence of $P(n)$ for various $\d=0,1,2$, and $3$.
}
 \label{fig:P}
\end{figure}

To give another support on our expectation, we will show another {\it density} of negative coefficients almost monotonically decrease as $n$ increases. we define the density with the window $[n, n+100]$ as
\begin{equation}\label{eq:den2}
P(n) \equiv \fr{\# \text{ of negative coefficients in } [n, n+100] }{50}.
\end{equation}
The $n$-dependence of $P(n)$ for various $\d$ is shown in Figure \ref{fig:P}.
One can immediately find that the density $P(n)$ for $\d=-2,-\fr{9}{5}, -\fr{8}{5}$ decays and finally vanishes as $n$ becomes large. Moreover, the decay rate decreases as we increase $\d$.
From this observation, we naturally expect that the density $P(n)$ for $\d=-\fr{7}{5}, -\fr{6}{5}$ also decays and vanishes at a particular value of $n$, even though we cannot find this vanishing of $P(n)$ in our figure because of the limitation of our machine power (i.e., the cut-off of $n$ at $n=1500$).
On the other hand, the $n$-dependence for $\d=-1$ is apparently different from that for other $\d$, that is, the $P(n)$ for $\d=-1$ shows a nearly constant value $P(n) \sim 1$ for all $n$.
For $\d=0,1,2,3$, one can find $P(n) \sim \fr{1}{2}$, which is consistent with our expectation that the sign of the coefficients frequently oscillate if $\d \in [-1,5]$. Here we exclude the point $\d=-1$ due to its specialty.

\begin{figure}[t]
\centering
  \includegraphics[width=6.0cm,clip]{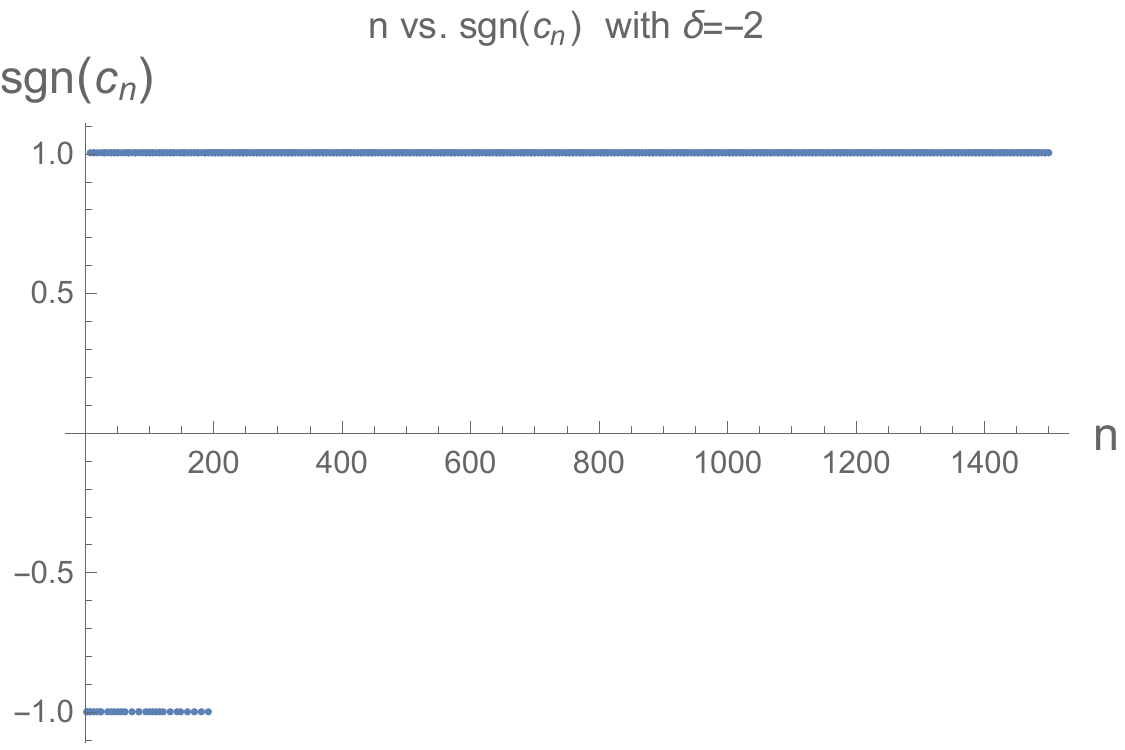}
  \includegraphics[width=6.0cm,clip]{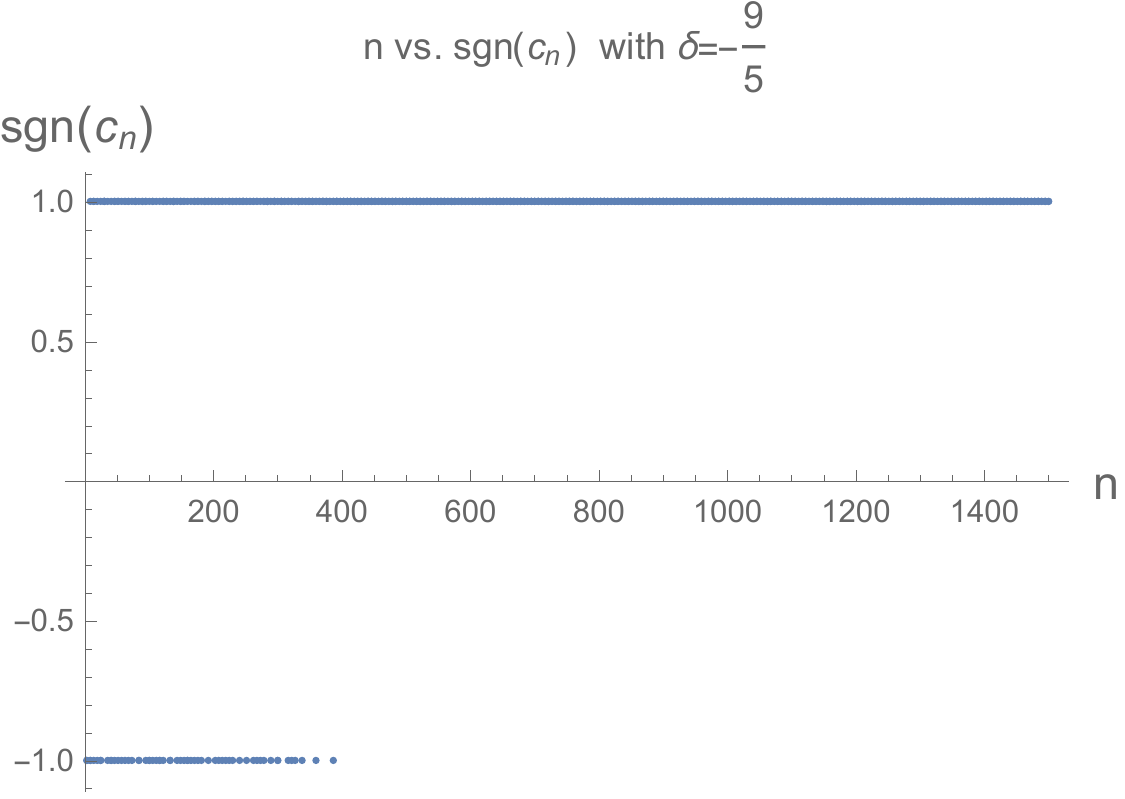}
  \includegraphics[width=6.0cm,clip]{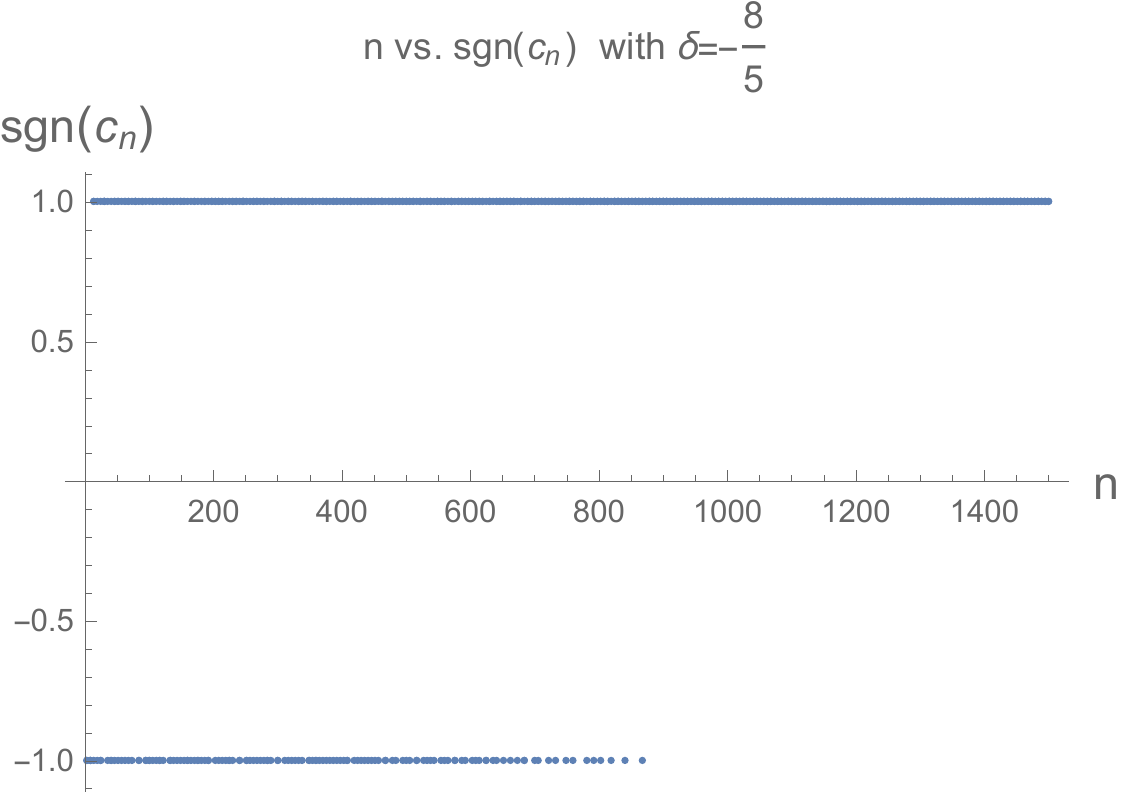}
  \includegraphics[width=6.0cm,clip]{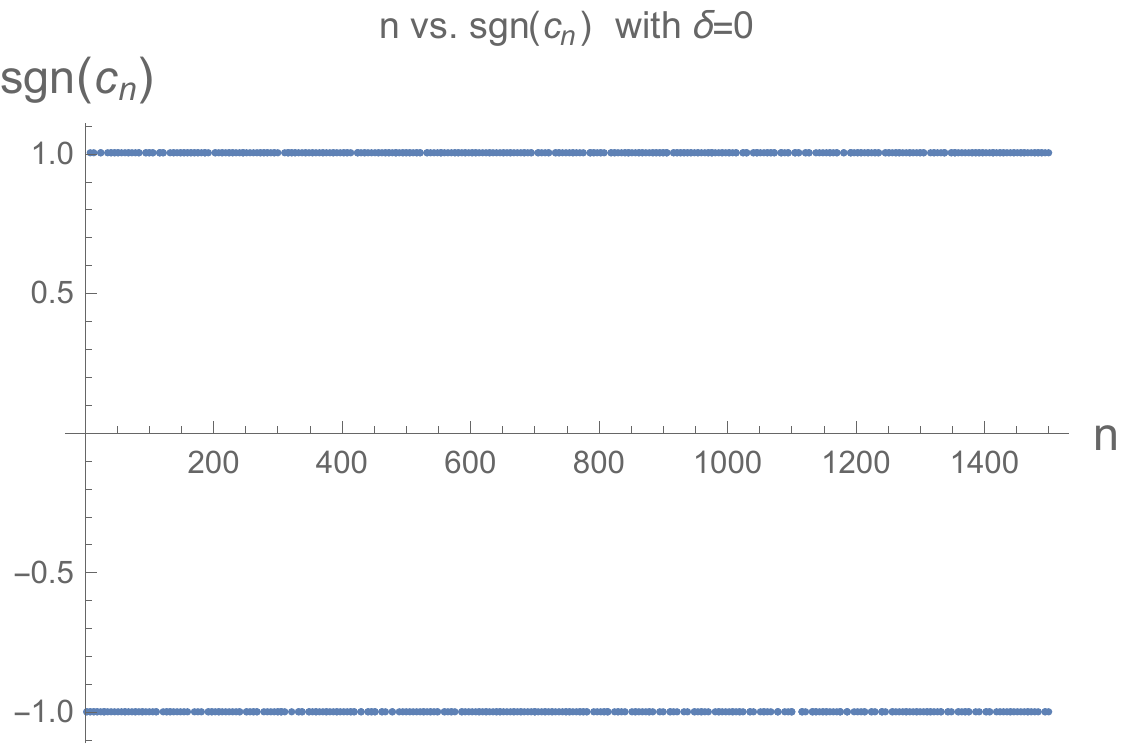}
 \caption{The $n$-dependence of $\sgn(c_n)$ for various $\d=-2,-\fr{9}{5}, -\fr{8}{5}$, and $0$ with $c=30$.}
 \label{fig:ex2}
\end{figure}

We summarize with the following conjectures,
\begin{itemize}
\item $c_n$ can be negative for all $n$ in the region $h_\O \in [\fr{c-1}{32}, \fr{c+5}{32}]$.
\item It is also possible that $c_n$ becomes negative in  $h_\O \notin [\fr{c-1}{32}, \fr{c+5}{32}]$, but the contributions to $H^{AA}_{BB}(q)$ (with $h_A=h_B$) from the negative coefficients might be negligible in the limit $q \to 1$ because we have $P(n) \ar{n \to \infty} 0$.

\end{itemize}

\begin{figure}[t]
\centering
  \includegraphics[width=6.0cm,clip]{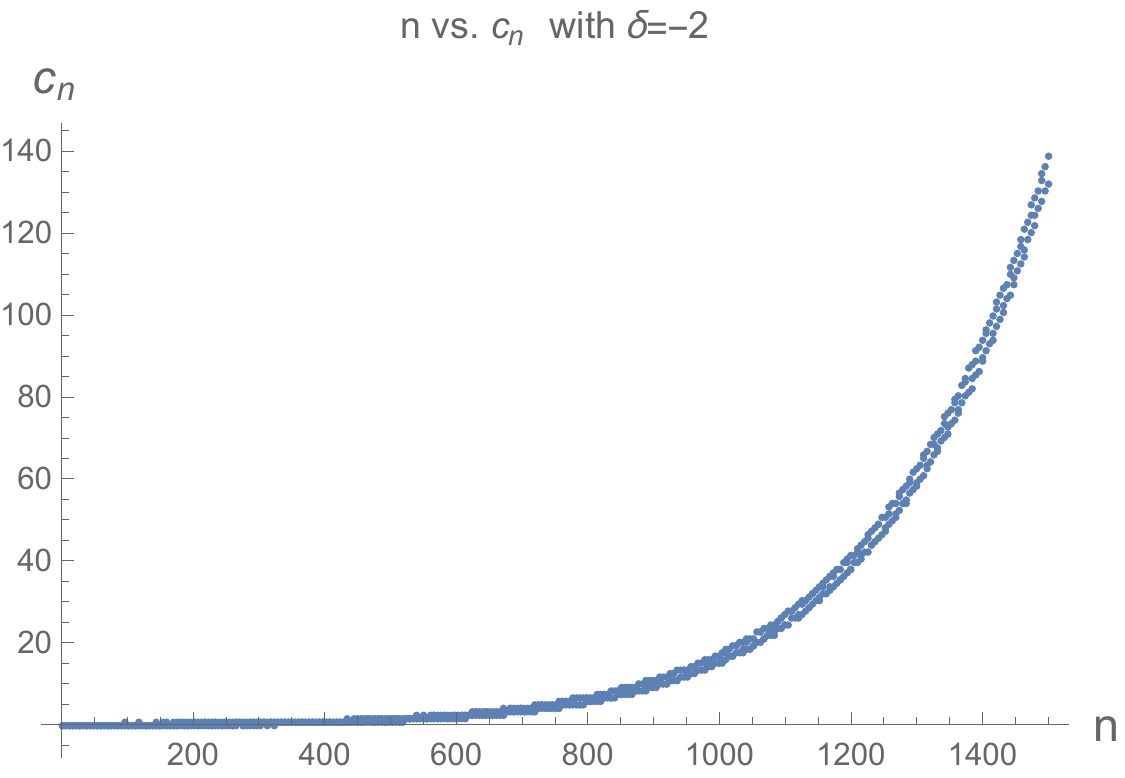}
  \includegraphics[width=6.0cm,clip]{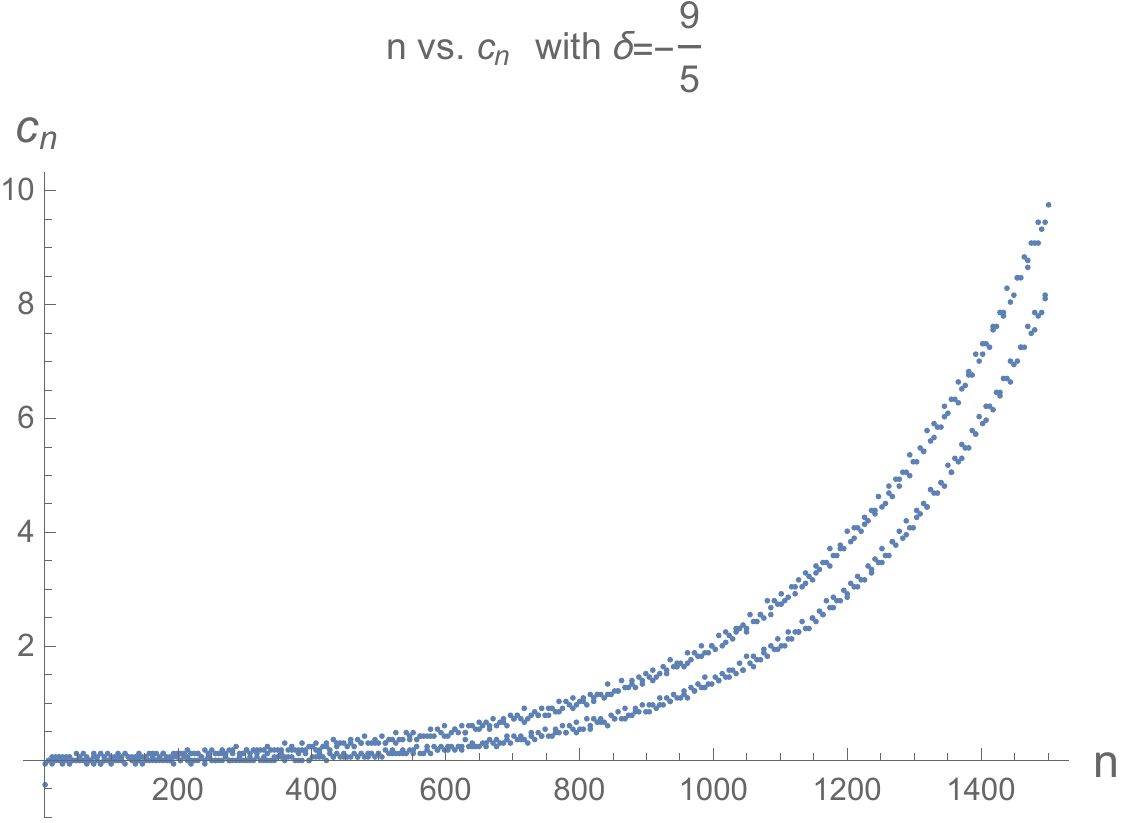}
  \includegraphics[width=6.0cm,clip]{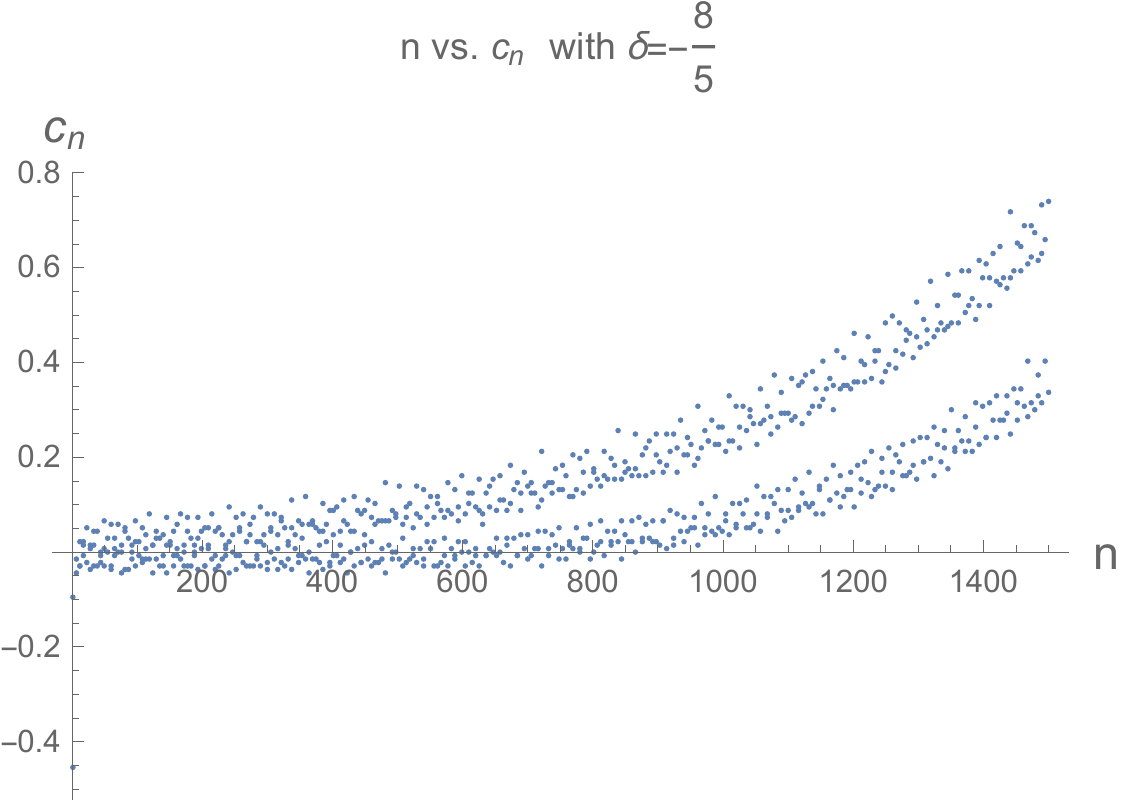}
  \includegraphics[width=6.0cm,clip]{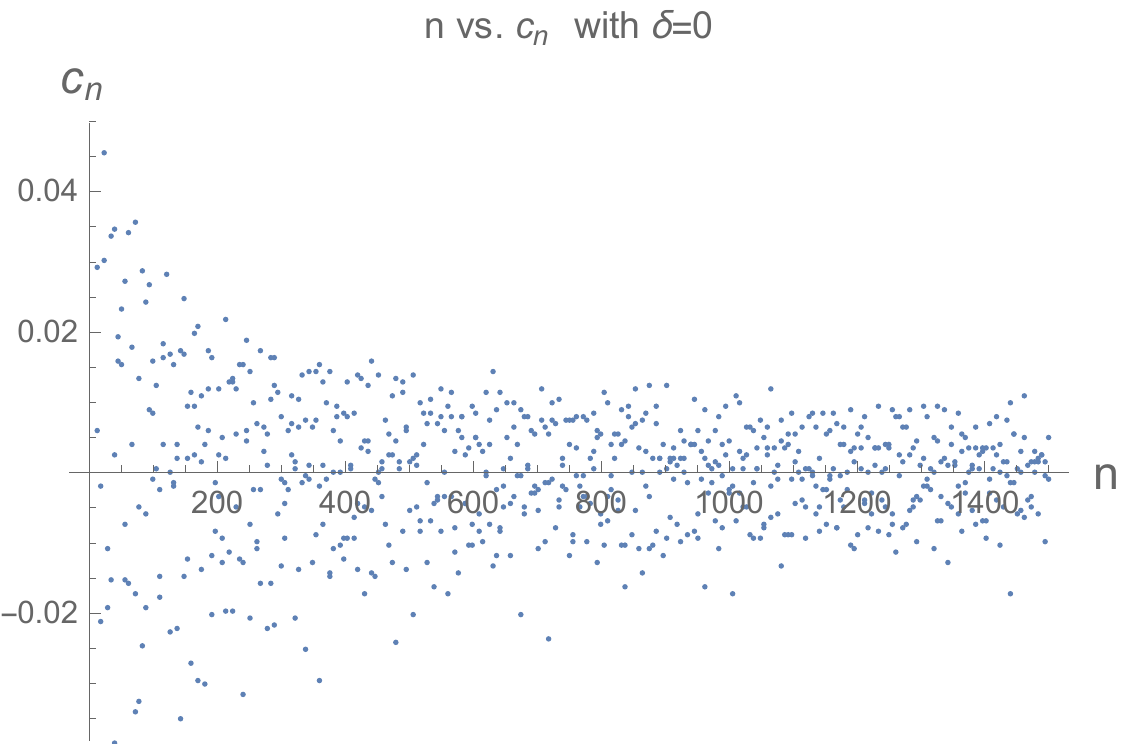}
 \caption{The $n$-dependence of $c_n$ for various $\d=-2,-\fr{9}{5}, -\fr{8}{5}$, and $0$ with $c=30$.}
 \label{fig:cn}
\end{figure}

One can observe the second conjecture explicitly in Figure \ref{fig:ex2}, where $P(n)$ for $\d=-2,-\fr{9}{5}, -\fr{8}{5}<-1$ vanishes in the limit $n \to \infty$ whereas the $\sgn(c_n)$ for $\d=0$ oscillates very fast.
In fact, this disappearing of the negative coefficients can be intuitively understood from the $n$-dependence of $c_n$, as shown in Figure \ref{fig:cn}. This plot naturally implies that the $c_n$ is roughly monotonically increasing in $n$. As a result, we find the disappearing of the negative coefficients for  $\d=-2,-\fr{9}{5}, -\fr{8}{5}$.
On the other hand, we cannot find such a growth for $\d=0$.

\subsection{Consequences from the positivity}

It turns out that for $h_A,h_B\notin \left(\frac{c-1}{32},\frac{c+5}{32}\right)$ and $h_A=h_B$, $c_n$ eventually becomes positive as $n$ is increased. Thus one can safely apply Tauberian theorems (similar theorem is used in \cite{Qiao:2017xif} in context of $0+1$D CFT). The asymptotic result has already been obtained in \cite{Brehm:2019pcx}. Here we will not repeat the method but will just point out how knowing $H(h,q\to1)$ gives immediately the growth of $c_n$. The numerics tells us that $c_n$ has growth eventually, in particular we have 
$$c_{n+1}-c_n>0\,.$$ This allows us to take to consider $$\frac{d}{dq}H(h,q)=\sum_{n} (c_{n}-c_{n-1})q^{n}$$

$\bullet$ Now one can apply Tauberian theorems of the following form:\\
 
\textit{Say, we have $F(q)=\sum a_n q^{n}$ with $a_n\geq 0$, then $\sum_{k=1}^{N}a_n \sim \int_{0}^{N} \mathcal{L}^{-1} \left[F(q\to 1)\right]$ where $\mathcal{L}^{-1}$ is the inverse Laplace transformation.}\\

In our case, we have $F(q)=\frac{d}{dq}H(h,q)$ and $\sum_{k=1}^{N}(c_{k}-c_{k-1})=c_{N}$. Similar arguments are also presented in appendix of \cite{Mukhametzhanov:2020swe}.

The above argument clearly shows that 
$$c_n \sim \mathcal{L}^{-1} \left[H(h,q\to 1)\right]\,.$$
Thus the naive procedure followed in \cite{Brehm:2019pcx} works.
We further remark that the asymptotics has been previously predicted numerically in \cite{kusuki1}. Here our approach is to take some input from the numerics and justify the rest of the analysis analytically. 

\subsection{Large $h$ asymptotics of $H(q)$} \label{subsec:lhp}

Here, we would like to find the asymptotics,
\begin{equation}\label{eq:H1}
H \pa{ h,e^{-\pi\sqrt{\frac{c-1}{6h}}} }  \ar {h \to \infty} ?.
\end{equation}
This quantity is relevant for the asymptotic estimate of the OPE coefficients. We naively expect this limit to converge to 1 from the well-known asymptotics with $q$ fixed,
\begin{equation}
H \pa{ h,q}   \ar{h \to \infty} 1, \ q\ \text{fixed}\,,
\end{equation}
but it should be mentioned that we are intersted in a simultaneous limit, as $h\to \infty$, we have $q\to1$. This is what makes the asymptotics nontrivial.
\begin{figure}[t]
 \begin{center}
  \includegraphics[width=12.0cm,clip]{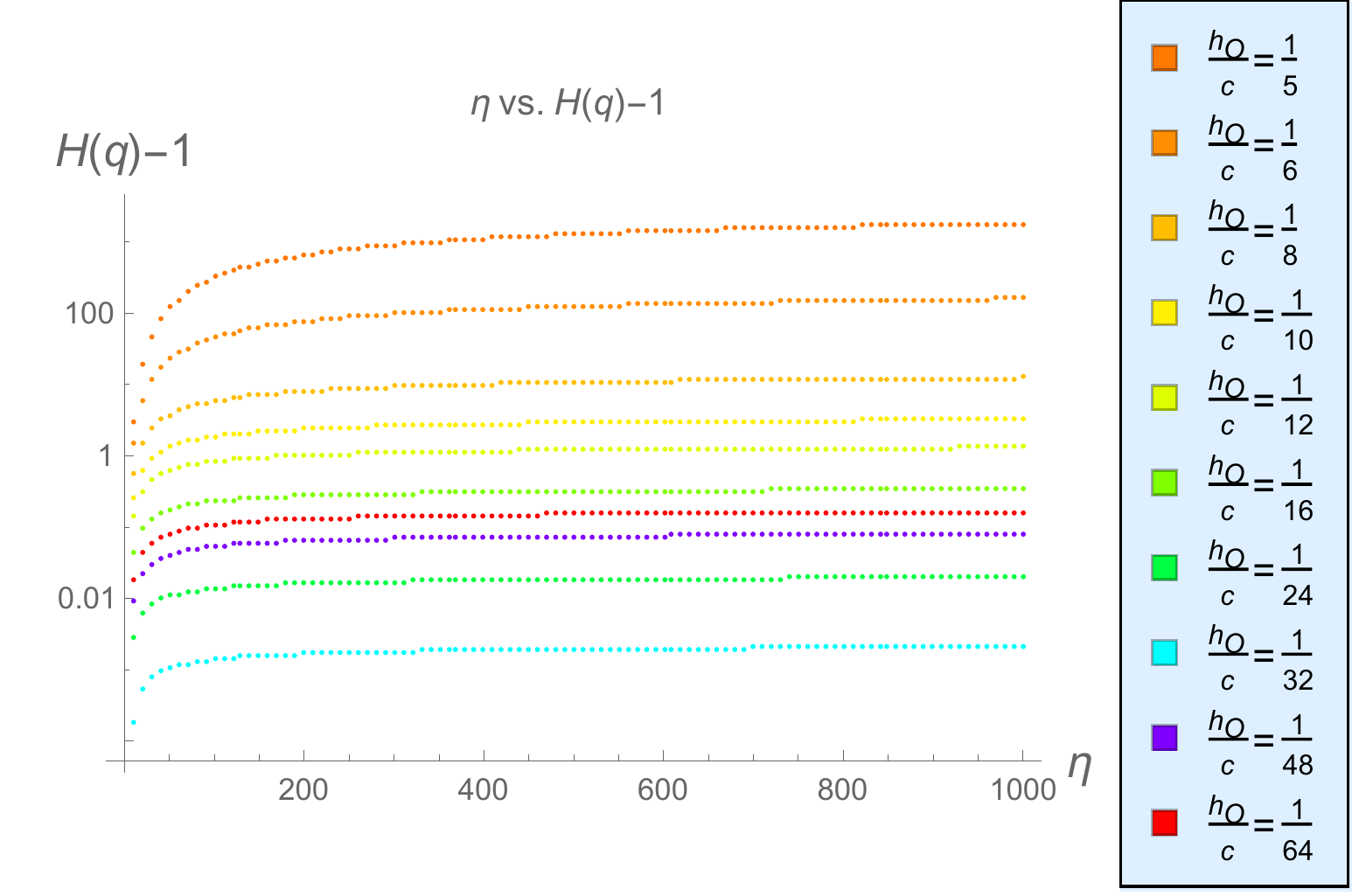}
 \end{center}
 \caption{Colored curves show the $h$-dependence of $H \pa{ h,e^{-\pi\sqrt{\frac{c-1}{6h}}} }-1$ with $h_\ca{O}=\fr{c}{6}, \fr{c}{8}, \fr{c}{10}, \fr{c}{12}, \fr{c}{32}, \fr{c}{64}$. For convenience, we introduce a normalized internal dimension $\eta=\fr{32}{c}h$. One can see that for any $h_\ca{O}$, the function $H$ converges to some positive values, which we want to show.}
 \label{fig:hpdep}
\end{figure}


\begin{figure}[t]
 \begin{center}
  \includegraphics[width=12.0cm,clip]{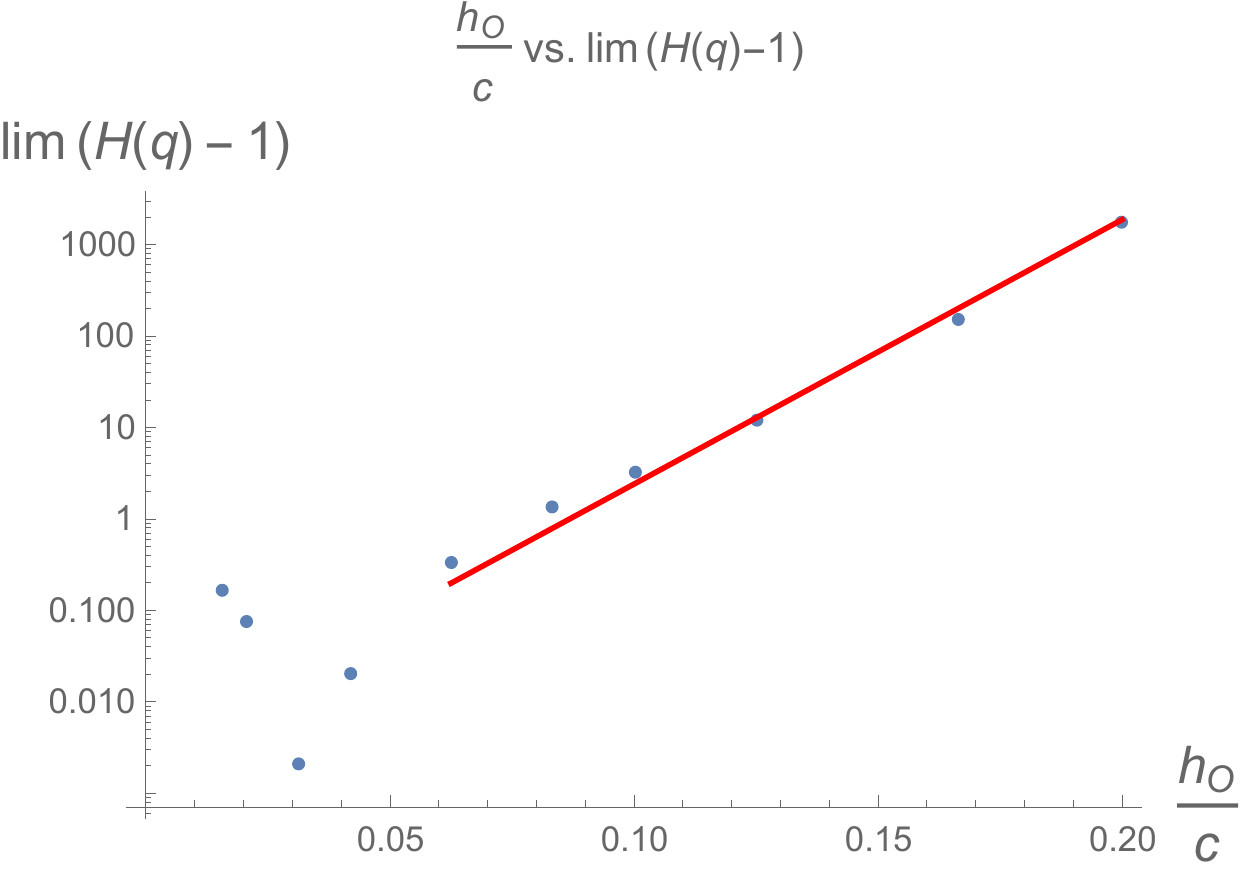}
 \end{center}
 \caption{ The $h_\ca{O}$-dependence of $H \pa{ h,e^{-\pi\sqrt{\frac{c-1}{6h}}} }-1$ at $h_p = 900$. One can find that the saturation value $H_0$ shows a linear dependence on $h_\ca{O}$, except for $h_\ca{O}=\fr{c}{32}$. This exception can be beautifully explained from the results in \cite{Cardona:2020cfy, Das:2020fhs}. The veritcal axis is in logscale.}
 \label{fig:hdep}
\end{figure}

We evaluated this asymptotics by the Zamolodchikov recursion relation up to $q^{1000}$. In Figure \ref{fig:hpdep}, we showed the $h$-dependence of the difference $H \pa{ h,e^{-\pi\sqrt{\frac{c-1}{6h}}} } - 1$ with different values of $h_\ca{O}/c$ for example, $h_\ca{O}=\fr{c}{6}, \fr{c}{8}, \fr{c}{10}, \fr{c}{12}, \fr{c}{32}, \fr{c}{64}$. One can obviously see that the difference does not converge to 0. Therefore, we can conclude that our simultaneous limit has non-trivial asymptotics, unlike the $h\to \infty$ with $q$ fixed.
 
Let us read off universal properties from the Figure \ref{fig:hpdep}.
We first would like to mention that the function $H(h,q)$ seems to converge to a positive number for any $h_\ca{O}$.
This observation justifies the technical assumption made in \eqref{assumption}. We also study the $h_\ca{O}$-dependence of the saturation value, denoted as $H_0$.
See Figure \ref{fig:hdep}, where we show the $h_\ca{O}$-dependence of $H \pa{ h,e^{-\pi\sqrt{\frac{c-1}{6h}}} }-1$ at $h_p = 900$. 
One finds numerically that the logarithmic saturation value $\log \,H_0$ behaves like a linear function of $h_\ca{O}$, except for $h_\ca{O}=\fr{c}{32}$.
This implies that the asymptotics can be expressed by
\begin{equation}
H \pa{ h,e^{-\pi\sqrt{\frac{c-1}{6h}}} }  \ar {h \to \infty} e^{\#h_\ca{O}}.
\end{equation}
We do not have any analytic proof of this formula.
It would be very interesting to show an analytic proof of this formula, but we have no proof at present.
We leave it as a future work.

The exception $h_\ca{O}=\fr{c}{32}$ can be explained from the results in \cite{Cardona:2020cfy, Das:2020fhs}.\footnote{This special value $\fr{c}{32}$ appears many times in the context of two-dimensional conformal bootstrap, for example,  \cite{Kusuki:2019gjs, Collier:2018exn}}.
According to \cite{Cardona:2020cfy, Das:2020fhs}, 
the leading and next-leading terms are written as
\begin{equation}\label{eq:next}
H(h,q) = 1 - \fr{1}{16 h}\pa{(c+1)-32h_{\ca{O}}}\pa{(c+5)-32h_{\ca{O}}}\pa{\fr{E_2(q)-1}{24}} + \ca{O}(1/h^2),
\end{equation}
where $E_2(q)$ is the Eisenstein series. 
This result shows that the large $h$ limit with fixed $q$ leads to the well-known result by Zamolodchikov, but in the large $h$ limit with $\beta=\pi\sqrt{\frac{c-1}{6h}}$, the next-leading cannot be neglected.
This is consistent with our findings in Figure \ref{fig:hpdep}, where we see the large correction to $H=1$.
Moreover, the equation (\ref{eq:next}) shows that the corrections would be small if $h_\ca{O}-\fr{c}{32} \ll 1$.
This result simply explains why the saturation value of $H$ (i.e $H_0$) becomes small if $h_\ca{O}=\fr{c}{32}$.

It should be mentioned that if one wants to consider the large $c$ limit, one cannot rely on the result (\ref{eq:next}).
One can immediately find that the large $c$ limit with $h_{\ca{O}}/c$ and $h/c$ fixed leads to a divergence of the next-leading term ($E_2(q) $ is of order $1$ in this case) unless $h_\ca{O} \sim c/32$. It means that we cannot commute two operations, the large $c$ limit and the large $h$ expansion (\ref{eq:next}) for this calculation. Nevertheless, in the large $c$ limit case, we can make use of another method.
In \cite{Kusuki2018}, the author discusses the validity region where we can approximate $H \simeq 1$ in the large $c$ limit by using the monodromy method \footnote{Note that this is the generalization of the validity region of the original Zamolodchikov monodromy method, $h\beta^2 \gg c$ \cite{Kusuki2018}.}.
\begin{equation}
\fr{1}{\beta^2} = \ca{O}\pa{h}.
\end{equation}
Our interest is $\beta=\pi\sqrt{\frac{c-1}{6h}}$, which satisfies the above validity condition.
As a result, we can conclude that if we first take the large $c$ limit with $h/c$ fixed, we obtain
\begin{equation}
H(h,e^{-\pi\sqrt{\frac{c-1}{6h}}}) = 1 + \ca{O}(1/c).
\end{equation}

\section{Asymptotics of $n$ point correlator in heavy states \& ETH}\label{Sec:npoint}
The $n$ point correlators in the heavy state are known to be very excellent proxy for the thermal correlators. The statement is true in some averaged sense. In this section, we revisit the statement, clarifying the notion of the averaging. Our statement is going to be for all CFTs and the optimal window would be order one. We expect that for ``chaotic'' CFTs, the window can be shrunk exponentially small in some large parameter ( $\sim e^{-\#c}$). The salient feature of this analysis is to show existence of an enigmatic regime where the heavy states don't dominate the canonical ensemble. This is similar in nature in the enigmatic regime found in \cite{HKS}.

In this section, we discuss $2$ point correlator and everything almost immediately generalizes to $n$ point correlator. We mostly focus on large central CFTs with sparse low lying spectrum. In large central charge limit, the universal behavior persists till $t\simeq \beta$, which is finite. Here $t$ is the Euclidean time.  The mathematical analysis can be done for finite central charge but then we have to let $\beta\to 0$ , as a result insertion time of $n$ operators has to go to $0$ as well.

\subsection{Tauberian theorem(s) for the infinite temperature torus correlator}

We start with an un-normalized torus two point correlator $\langle \mathcal{O}(t)\mathcal{O}(0)\rangle_{\beta}$. Here we use $t$ to denote the Euclidean time and $\beta$ is the inverse temperature.  Our objective in this section is to have a rigorous estimate of the following quantity in the $\Delta\to\infty$ limit:
\begin{equation}
\begin{aligned}
\mathcal{B} (\Delta) &=\int_{\Delta-\delta}^{\Delta+\delta}\text{d}\Delta'\ \left(\sum_{\Delta_{w}=\Delta_{ph}}\langle w|\mathcal{O}(t)\mathcal{O}(0)|w\rangle \right) \delta(\Delta'-\Delta_{ph})\\
&=\int_{\Delta-\delta}^{\Delta+\delta}\text{d}\Delta'\ b(\Delta^\prime)
\end{aligned} 
\end{equation} 
where we have defined
\begin{align}
b(\Delta^\prime)= \left(\sum_{\Delta_w=\Delta_{ph}} \langle w|\mathcal{O}(t,0)\mathcal{O}(0)|w\rangle\right)\delta(\Delta^\prime-\Delta_{ph})
\end{align}

The rigorous estimation will provide us with a precise sense in which a two point correlator in heavy state can mimic a thermal state. We note that $b(\Delta')$ appears in the following expansion of unnormalized two point correlator.  
\begin{equation}\label{expansion}
\begin{aligned}
\langle \mathcal{O}(t,0)\mathcal{O}(0)\rangle_{\beta}= \sum_{w} \langle w|\mathcal{O}(t,0)\mathcal{O}(0)|w\rangle e^{-\beta(\Delta-c/12)}=\int_{0}^{\infty} \text{d}\Delta^\prime b(\Delta^\prime)e^{-\beta(\Delta-c/12)}
\end{aligned}
\end{equation}
Thus the idea is to extract the $\beta\to 0$ limit of the torus two point correlator and deduce the asymptotic behaviour of $b(\Delta')$.\\ 

$\bullet$ \textbf{Large central charge}:\\

In what follows, we will be considering large central charge limit and let 
\begin{align}
\Delta=c\left(\frac{1}{12}+\epsilon\right)\,,
\end{align}
and in this limit, we will be estimating $\mathcal{B}(\Delta)$. We start with the basic inequality: 
\begin{align}
\phi_-(\Delta^\prime) \leq \Theta( \Delta^\prime) \leq \phi_+(\Delta^\prime) \,.
\end{align}

For brevity, we will write the argument for the upper bound (the lower bounds works out in a similar manner) 
From the above, we write the following
\begin{equation}
\begin{aligned}
\int_{\Delta-\delta}^{\Delta+\delta} \text{d}\Delta^\prime\  b(\Delta^\prime)
\leq  e^{\beta(\Delta+\delta)}\int_{0}^{\infty} \text{d}\Delta^\prime\ b(\Delta^\prime) \phi_+(\Delta^\prime) e^{-\beta\Delta'} \,,
\end{aligned}
\end{equation}

This leads to

\begin{equation}
\begin{aligned}\label{ineq1}
\int_{\Delta-\delta}^{\Delta+\delta} \text{d}\Delta^\prime\ b(\Delta^\prime)\leq  e^{\beta(\Delta+\delta-c/12)}\int_{-\Lambda}^{\Lambda} \text{d}\Omega\ \langle \mathcal{O}(t,\phi)\mathcal{O}(0)\rangle_{\beta+\imath\Omega}\  \widehat{\phi}_+(\Omega)
\end{aligned}
\end{equation}

Modular covariance tells us that
\begin{align}
\langle \mathcal{O}(t,\phi)\mathcal{O}(0)\rangle_{\beta}=  \left(\frac{2\pi}{\beta}\right)^{2\Delta_{\mathcal{O}}} \bigg\langle \mathcal{O}\left(\frac{2\pi\phi}{\beta},\frac{2\pi t}{\beta}\right)\mathcal{O}(0)\bigg\rangle_{\frac{4\pi^2}{\beta}}
\end{align}

We plug that in \eqref{ineq1} to obtain
\begin{equation}
\begin{aligned}
&\int_{\Delta-\delta}^{\Delta+\delta} \text{d}\Delta^\prime\ b(\Delta^\prime)\\
&\leq  e^{\beta(\Delta+\delta-c/12)}\int_{-\infty}^{\infty} \text{d}\Omega\ e^{-\imath\Omega c/12} \left(\frac{2\pi}{\beta}\right)^{2\Delta_{\mathcal{O}}} \bigg\langle \mathcal{O}\left(0,\frac{2\pi t}{\beta+\imath\Omega}\right)\mathcal{O}(0)\bigg\rangle_{\frac{4\pi^2}{\beta+\imath\Omega}} \phi_+(\Omega)
\end{aligned}
\end{equation}

Now we separate out the light part and heavy part by looking at the expansion \eqref{expansion} and defining:
\begin{align}
\langle \mathcal{O}(t,\phi)\mathcal{O}(0)\rangle_{\beta}^{L}&= \sum_{w, \Delta_w<c/12} \langle w|\mathcal{O}(t,\phi)\mathcal{O}(0)|w\rangle e^{-\beta(\Delta-c/12)}\\
\langle \mathcal{O}(t,\phi)\mathcal{O}(0)\rangle_{\beta}^{H}&= \sum_{w, \Delta_w>c/12} \langle w|\mathcal{O}(t,\phi)\mathcal{O}(0)|w\rangle e^{-\beta(\Delta-c/12)}
\end{align}

Now use large $c$ vacuum dominance 
\begin{align}\label{vacuum}
\left(\frac{2\pi}{\beta+\imath\Omega}\right)^{2\Delta_{\mathcal{O}}}\left\langle \mathcal{O}\left(0,\frac{2\pi t}{\beta+\imath\Omega}\right)\mathcal{O}(0)\right\rangle_{\frac{4\pi^2}{\beta+\imath\Omega}}^{L}= e^{\frac{\pi^2c}{3(\beta+\imath\Omega)}} \left(\frac{2\pi}{\beta+\imath\Omega}\right)^{2\Delta_{\mathcal{O}}}\left[\sin\left(\frac{\pi t}{\beta+\imath\Omega}\right)\right]^{-2\Delta_{\mathcal{O}}}
\end{align}

The contribution from the light part is dominated by $\Omega=0$ and we have
\begin{equation}
\begin{aligned}
&e^{\beta(\Delta+\delta-c/12)}\int_{-\infty}^{\infty} \text{d}\Omega\ e^{-\imath\Omega c/12} \left(\frac{2\pi}{\beta(\Omega)}\right)^{2\Delta_{\mathcal{O}}} \bigg\langle \mathcal{O}\left(0,\frac{2\pi t}{\beta+\imath\Omega}\right)\mathcal{O}(0)\bigg\rangle^{L}_{\frac{4\pi^2}{\beta+\imath\Omega}} \phi_+(\Omega)\\
&\underset{c\to \infty} {\simeq } \sqrt{\frac{3}{\pi c}}\beta^{3/2} e^{\frac{\pi^2c}{3\beta}+\beta(\Delta+\delta-c/12)} \left(\frac{2\pi}{\beta}\right)^{2\Delta_{\mathcal{O}}}\left[\sin\left(\frac{\pi t}{\beta}\right)\right]^{-2\Delta_{\mathcal{O}}}+O(1)
\end{aligned}
\end{equation}
where $\beta(\Omega)=\beta+\imath\Omega$.

Next we will show that the heavy part is suppressed compared to the light part. We can proceed as  we did for the $4$ point correlator and bound the absolute value: 

\begin{equation}
 \bigg|\bigg\langle \mathcal{O}\left(0,\frac{2\pi t}{\beta+\imath\Omega}\right)\mathcal{O}(0)\bigg\rangle_{\frac{4\pi^2}{\beta+\imath\Omega}}^H\bigg| \leq \sum_{H}\bigg|\langle H|\mathcal{O}\left(0,\frac{2\pi t}{\beta+\imath\Omega}\right)\mathcal{O}(0)|H\rangle\bigg| e^{-\frac{4\pi^2\beta}{\beta^2+\Omega^2}(\Delta_H-c/12)}
\end{equation}

Thus the absolute value of the $\Omega$ integral coming from the heavy part can be estimated as
$$e^{\beta(\Delta\pm\delta-c/12)}\int_{-\Lambda}^{\Lambda} d\Omega\ \left(\frac{2\pi}{\beta(\Omega)}\right)^{2\Delta_{\mathcal{O}}}\left(\sum_{H}\bigg|\langle H|\mathcal{O}\left(0,\frac{2\pi t}{\beta(\Omega)}\right)\mathcal{O}(0)|H\rangle\bigg| e^{-\frac{4\pi^2\beta}{\beta^2+\Omega^2}(\Delta_H-c/12)}\right)|\widehat{\phi}_{\pm}(\Omega)|\,,$$
 is dominated $\Omega\simeq O(1)$ number. Thus we go to the dual channel and estimate the contribution. For suppression, one needs to choose $$\Lambda < 2\pi\sqrt{1-\frac{1}{12\epsilon}}\,.$$
This requires $\epsilon>1/12$ and $\Delta>c/6$. Combining everything and setting $\beta=\pi\sqrt{\frac{1}{3\epsilon}}$, we have
\begin{equation} 
\begin{aligned}
&\log\bigg[\int_{\Delta-\delta}^{\Delta+\delta} \text{d}\Delta' \left(\sum_{\Delta_w=\Delta_{ph}} \langle w| \mathcal{O}(t,0)\mathcal{O}(0) |w \rangle\right) \bigg]\delta(\Delta'-\Delta_{ph})\\ &
\underset{c\to\infty}{=} 2\pi c \sqrt{\frac{\epsilon}{3}} + \frac{1}{2} \log (c) -2\Delta_{\mathcal{O}}\log\left[\sin\left(t \sqrt{3\epsilon}\right)\right] +O(1)
\end{aligned} 
\end{equation}

One can keep track of the order one number as in \cite{baur} and arrive at following bounds:
\begin{equation}
\begin{aligned}\label{holo}
&\ \left(\widehat{\phi}_{-}(0)+\ell\right) e^{-\beta\delta}\rho_0(\Delta)\left(\frac{2\pi}{\beta}\right)^{2\Delta_{\mathcal{O}}}\left[\sin\left(\frac{\pi t}{\beta}\right)\right]^{-2\Delta_{\mathcal{O}}}\\
&\leq \mathcal{B}_{-}\leq\mathcal{B}_+ \\
&\leq \left(\widehat{\phi}_{-}(0)+\ell\right)e^{\beta\delta}\rho_0(\Delta)\left(\frac{2\pi}{\beta}\right)^{2\Delta_{\mathcal{O}}}\left[\sin\left(\frac{\pi t}{\beta}\right)\right]^{-2\Delta_{\mathcal{O}}}
\end{aligned}
\end{equation}
The quantity $\ell$ signifies the positive contribution coming from the sparse low lying states. Now we need to use Beurling-Selberg functions \cite{Mukhametzhanov:2020swe} maintaining the constraint $\Lambda < 2\pi\sqrt{1-\frac{1}{12\epsilon}}=2\pi\sqrt{1-\frac{\beta^2}{4\pi^2}}$. This leads to 
\begin{equation}
\widehat{\phi}_{\pm}(0)=\left(2\delta\pm\frac{2\pi}{\Lambda}\right)=\left(2\delta\pm\left(1-\frac{\beta^2}{4\pi^2}\right)^{-1/2}\right)
\end{equation}

Now the number of states within the bin in large $c$ limit is given by
\begin{equation}
\log\bigg[\int_{\Delta-\delta}^{\Delta+\delta} \text{d}\Delta' \rho(\Delta')\bigg]
\underset{c\to\infty}{=} 2\pi c \sqrt{\frac{\epsilon}{3}} + \frac{1}{2} \log (c) +O(1)
\end{equation}

This implies the average value of two point correlator in heavy state is given by thermal correlator at $\beta_{bh}=\pi\sqrt{\frac{c}{3(\Delta-c/12)}}$
\begin{equation}
\begin{aligned}
 \langle H| \mathcal{O}(t,0)\mathcal{O}(0) |H \rangle_{\text{average}} \underset{c\to\infty}{=} \left[\sin\left(t \sqrt{3\epsilon}\right)\right]^{-2\Delta_{\mathcal{O}}}=\langle  \mathcal{O}(t,0)\mathcal{O}(0) \rangle_{\beta_{bh}}\\
 \end{aligned}
\end{equation}

Another way of phrasing the result is following:
\begin{equation}
 \langle H| \mathcal{O}(t,0)\mathcal{O}(0) |H \rangle_{\text{average}} \simeq \text{Tr} \left( e^{-\beta H}\mathcal{O}(t,0)\mathcal{O}(0) \right)+ O(1)
\end{equation}

In fact, by smilar arguments one can show that
\begin{equation}
 \langle H| \mathcal{O}(t_1,0)\mathcal{O}(t_2,0)\cdots \mathcal{O}(t_n,0)|H \rangle_{\text{average}} \simeq \text{Tr} \left( e^{-\beta H}\mathcal{O}(t_1,0)\mathcal{O}(t_2,0)\cdots \mathcal{O}(t_n,0)\right)+ O(1)
\end{equation}
as long as $n$ is finite and $\Delta_H>c/6$. The statement we have just proved is true for Euclidean time separated correlators. The are technical subtleties in analytically continuing the result in Lorentzian time and make more direct connection with the results of \cite{Anous:2019yku}. This happens precisely because of the presence of averaging window. Recall the size of averaging window is controlled by the heavy states in the dual channel. In particular, we need the tail to be suppresed. Analytically continuing to the Lorentzian time might spoil this suppression. It is worth investigating how to make the continuation very precise. Something along the line of proposal made in \cite{deBoer:2019kyr} might be useful in this regard. We conclude this section by making a brief statement for the finite central charge case. 

$\bullet$ \textbf{Finite central charge:} For finite central charge, one can repeat the above analysis. Now we have
\begin{equation}
\beta=\pi\sqrt{\frac{c}{3\Delta}}\to 0\,,\ \frac{t}{\beta}=\text{constant}\,.
\end{equation}
Effectively, we are looking at the correlator in $t\to 0$ limit while keeping $t/\beta$ finite. 

In this limit, we obtain
\begin{equation}
\begin{aligned}
&2\pi\widehat{\phi}_{-}(0)\left(\frac{c}{48\Delta^3}\right)^{1/4} \exp\left[2\pi\sqrt{\frac{c\Delta}{3}}\right] \left(\frac{2\pi}{\beta}\right)^{2\Delta_{\mathcal{O}}}\left[\sin\left(\frac{\pi t}{\beta}\right)\right]^{-2\Delta_{\mathcal{O}}}\\
&\leq \int_{\Delta-\delta}^{\Delta+\delta}d\Delta'\  \langle H| \mathcal{O}(t,0)\mathcal{O}(0) |H \rangle \leq\\
& 2\pi\widehat{\phi}_{+}(0)\left(\frac{c}{48\Delta^3}\right)^{1/4} \exp\left[2\pi\sqrt{\frac{c\Delta}{3}}\right]\left(\frac{2\pi}{\beta}\right)^{2\Delta_{\mathcal{O}}}\left[\sin\left(\frac{\pi t}{\beta}\right)\right]^{-2\Delta_{\mathcal{O}}}
\end{aligned}
\end{equation}

Using Selberg-Beurling functions, we obtain $2\pi\widehat{\phi}_{\pm}=2\delta\pm 1$. Thus the average two point function $\widehat{\mathcal{O}\mathcal{O}}\equiv \frac{\int_{\Delta-\delta}^{\Delta+\delta}d\Delta'\  \langle H| \mathcal{O}(t,0)\mathcal{O}(0) |H \rangle}{\int_{\Delta-\delta}^{\Delta+\delta}d\Delta'\  \rho(\Delta')} $ will be given by
\begin{equation}
\begin{aligned}
&\frac{2\delta-1}{2\delta+1}\left(\frac{2\pi}{\beta}\right)^{2\Delta_{\mathcal{O}}}\left[\sin\left(\frac{\pi t}{\beta}\right)\right]^{-2\Delta_{\mathcal{O}}}\leq\ \widehat{\mathcal{O}\mathcal{O}} \ \leq\frac{2\delta+1}{2\delta-1}\left(\frac{2\pi}{\beta}\right)^{2\Delta_{\mathcal{O}}}\left[\sin\left(\frac{\pi t}{\beta}\right)\right]^{-2\Delta_{\mathcal{O}}}\,,
\end{aligned} 
\end{equation}
where it is to be understood that $\beta$ is a function of $\Delta$. As $\Delta\to\infty$, $\beta\to 0$ and we are keeping $t/\beta$ fixed. 

The notion of optimal asymptotic spectral gap is bit trivial in this context if we allow for any operator $\mathcal{O}$. Then one can choose $\mathcal{O}$ to be identity,  and then the analysis precisely become the analysis for density of states. This why we arrive at same bound of asymptotic spectral gap as we have obtained in Cardy analysis \cite{baur, Ganguly:2019ksp,Mukhametzhanov:2020swe}.

\section*{Acknowledgements}
SP thanks Baur Mukhametzhanov for numerous stimulating generic discussions related to Tauberian theorems.  SP acknowledges the support from Ambrose Monell Foundation and DOE grant DE-SC0009988. DD would like to acknowledge the support provided by the Max Planck Partner Group grant MAXPLA/PHY/2018577. YK is supported by Grant-in-Aid for JSPS Fellows No. 18J22495.
{\appendix

\section{Simultaneous limit for $H(h,q)$: motivating the technical assumption} \label{app:assum}
We will motivate the assumption \eqref{assumption} in two steps. In order to do that, we need to understand why the simultaneous limit ($\beta_{L,R}\to0$, where $\beta_{L}=\pi\sqrt{\frac{c-1}{6h}}$ and $\beta_{R}=\pi\sqrt{\frac{c-1}{6\bar h}}$) is important. This will also set the stage for motivating our assumption on $H(h,q)$. Until now, we have set the length of the spatial circle of the torus to be $L=2\pi$ and let $\beta_{L,R}\to 0$. In CFT, using scale invariance, this is same as fixing $\beta_{L,R}$ and letting $L\to\infty$. So we restore $L$ for now and rewrite \eqref{pexp} as
\begin{equation}
p(\beta_L/L,\beta_R'/L)=\sum_{h', \bar{h}'}f^{2}_{\mathcal{O}\mathcal{O}O'}e^{-\frac{2\pi\beta_L}{2L}\left(h'-\frac{c-1}{24}\right)-\frac{2\pi\beta_R}{2L}\left(\bar h'-\frac{c-1}{24}\right)}H(h',q_L)H(\bar{h}',q_R)
\end{equation}
We define $E_{L}=\frac{2\pi}{L}(h-C/24)$ and $E_{R}=\frac{2\pi}{L}(\bar h-C/24)$, where $C=c-1$. In $L\to\infty$ limit, the variable $E_{L,R}$ become continous. Thus we can write
\begin{equation}
p(\beta_L/L,\beta_R'/L)=\int dE_L\ dE_R\ a(E_L,E_R)\ e^{-\frac{\beta_LE_L+\beta_RE_R}{2}}H(E_L,q_L)H(E_R,q_R)\,,
\end{equation}
and recast \eqref{modtrafo} as
\begin{equation}
\begin{aligned}
&\int dE_L\ dE_R\ a(E_L,E_R)\ e^{-\frac{\beta_LE_L+\beta_RE_R}{2}}H(E_L,q_L)H(E_R,q_R)\\
&\underset{\beta/L\to0}{=}\left(\frac{L^2}{\beta_L\beta_R}\right)^{\nu/2+1/4}\exp\left[\frac{\pi LC}{24\beta_L}+\frac{\pi LC}{24\beta_R}\right]
\end{aligned}
\end{equation}
So we make the following assumption $H(E_L,q_L),H(E_R,q_R)$ does not grow or fall exponentially in $E_L$ or $E_R$ in $q_{L,R}\to 1$ limit (or to rephrase in the neighborhood of $q=1$). Thus we can forget about $H$, still determine the saddle of the integral as a function of $\beta$ and $L$. This automatically provides us the saddle value and leading behavior of $a(E_L,E_R)$
\begin{equation}
\begin{aligned}
\beta_{L,R}&=\frac{1}{2}\sqrt{\frac{\pi CL}{3E_{L,R*}}}\\
\log a(E_{L*},E_{R*})&\simeq \sqrt{\frac{\pi CLE_L}{3}}+\sqrt{\frac{\pi CLE_R}{3}}+ \text{Errors}
\end{aligned}
\end{equation}
Finite $\beta_{L,R}$ implies that we are looking at finite energy density states. Even though this method provides us with leading behavior of $a(E_{L*},E_{R*})$, we can not immediately turn this into a function of $h_{L}$ and $h_{R}$, since there can be multiple states such that in $L\to\infty$ limit, $\frac{h_{L,R}}{L} \to E_{L,R}$. In short, we do not have any control over averaging window. Neither do we have any control over subleading corrections. Now we make the second step towards our assumption stated earlier
\begin{equation}
\lim_{L\to\infty} H(E_{L*},q_{L} \to 1)= H+ \text{error terms} 
\end{equation}
which can be equivalently written as (Recall we can exhange $\beta$ and $L$)
\begin{equation}
H(h \to\infty, q_{L} \to 1)= H + \text{error terms} \,,\  H>0\,\ \& \ q_{L}=e^{-\pi\sqrt{\frac{(c-1)h}{6}}}
\end{equation}
where $H$ is a constant and a similar equation holds for the right movers. In fact, we will need a slightly stronger assumtion:
\begin{equation}
H(h \to\infty, |q_{L}| \to 1)= H + \text{error terms}\,, \ H>0\,\ \&\  |q_{L}|=e^{-\pi\sqrt{\frac{(c-1)h}{6}}}
\end{equation}
This is the precisely the technical assumption that we have made in eq\eqref{assumption}.
}

{\bibliographystyle{bibstyle2017}
\bibliography{refs}
\hypersetup{urlcolor=RoyalBlue!60!black}
}

\end{document}